\documentclass[superscriptaddress,amsmath, nofootinbib]{revtex4}

\usepackage{amsfonts}

\usepackage{graphicx}

\usepackage[latin1]{inputenc}

\usepackage{amsmath}

\usepackage{amssymb}

\begin{document}



\title{Ultra-relativistic spinning particle and a rotating body in external fields}
\author{Alexei A. Deriglazov }
\email{alexei.deriglazov@ufjf.edu.br} \affiliation{Departamento de Matem\'atica, ICE, Universidade Federal de Juiz de
Fora, MG, Brazil} \affiliation{Laboratory of Mathematical Physics, Tomsk Polytechnic University, 634050 Tomsk, Lenin
Ave. 30, Russian Federation}

\author{Walberto Guzm\'an Ram\'irez }
\email{wguzman@cbpf.br} \affiliation{Departamento de Matem\'atica, ICE, Universidade Federal de Juiz de Fora, MG,
Brazil}

\date{\today}


\begin{abstract}
We use the vector model of spinning particle to analyze the influence of spin-field coupling
on the particle's trajectory in ultra-relativistic regime. The Lagrangian with minimal spin-gravity interaction yields
the equations equivalent to the Mathisson-Papapetrou-Tulczyjew-Dixon (MPTD) equations of a rotating body. We show that
they have unsatisfactory behavior in the ultra-relativistic limit. In particular, three-dimensional acceleration of the
particle increases with velocity and becomes infinite in the ultra-relativistic limit. The reason is that in the equation for trajectory
emerges the term which can be thought as an effective metric generated by the minimal spin-gravity coupling. Therefore
we examine the non-minimal interaction through the gravimagnetic moment $\kappa$, and show that the theory with
$\kappa=1$ is free of the problems detected in MPTD-equations. Hence the non-minimally interacting theory seem more
promising candidate for description of a relativistic rotating body in general relativity.

The Lagrangian for the particle in an arbitrary electromagnetic field in Minkowski space leads to generalized Frenkel
and Bargmann-Michel-Telegdi equations. The particle with magnetic moment in electromagnetic field and the particle with
gravimagnetic moment in gravitational field have very similar structure of equations of motion. In particular, the
spin-electromagnetic coupling also produces an effective metric for the particle with anomalous magnetic moment. If we
use the usual special-relativity notions for time and distance, then the critical speed, which the particle cannot
exceed during its evolution in electromagnetic field, is different from the speed of light. This can be corrected
assuming that the three-dimensional geometry should be defined with respect to the effective metric.
\end{abstract}

\maketitle 





\section{Introduction}\label{sec1}
The problem of a covariant description of rotational degrees of freedom has a long and fascinating
history\cite{Trautman2002}-\cite{corben:1968}. Equations of motion of a rotating body in curved background formulated usually
in the multipole approach to description of the body, see \cite{Trautman2002} for the review. The first results were
reported by Mathisson \cite{Mathisson:1937zz} and Papapetrou \cite{Papapetrou:1951pa}. They assumed that the structure
of test body can be described by a set of multipoles and have taken the approximation which involves only first two
terms (the pole-dipole approximation). The equations are then derived by integration of conservation law for the
energy-momentum tensor, $T^{\mu\nu}{}_{;\mu}=0$. A manifestly covariant equations were formulated by Tulczyjew
\cite{Tulc} and Dixon \cite{Dixon1964, Dixon1965}. In the current literature they usually appear in the form given by
Dixon (the equations (6.31)-(6.33) in \cite{Dixon1964}), we will refer them as  Mathisson-Papapetrou-Tulczyjew-Dixon
(MPTD) equations. They are widely used now to account spin effects in compact binaries and rotating black holes, see
\cite{Pomeranskii1998, Will2014, Balakin2015} and references therein.

Concerning the equations of spinning particle in electromagnetic field, may be the best candidates are those of Frenkel
\cite{Frenkel, Frenkel2} and Bargmann, Michel and Telegdi (BMT) \cite{BMT}. Here the strong restriction on possible
form of semiclassical equations is that the reasonable model should be in correspondence with the Dirac equation. In
this regard the vector model of spin (see below) is of interest because it yields the Frenkel equations at the
classical level, and implies the Dirac equation after canonical quantization \cite{DPM2}.

In this work we study behavior of a particle governed by these equations (as well as by some of their generalizations)
in the ultra-relativistic regime. To avoid the ambiguities in the passage from Lagrangian to Hamiltonian description
and vice versa, and in the choice of possible form of interaction, we start in each case from an appropriate
variational problem. The vector models of spin provide one possible way to achieve this. In these models, the basic
variables\footnote{For early attempts to build a vector model, see the review \cite{f}.} in spin-sector are
non-Grassmann vector $\omega^\mu$ and its conjugated momentum $\pi_\mu$. The spin-tensor is a composite quantity
constructed from these variables, $S^{\mu\nu}=2(\omega^\mu\pi^\nu-\omega^\nu\pi^\mu)$. To have a theory with right
number of physical degrees of freedom for the spin, certain constraints on the eight basic variables should follow from
the variational problem. It should be noted that even for the free theory in flat space, search for the variational
problem represents rather non trivial task (for the earlier attempts, see \cite{mukunda1982} and the review \cite{f}).

To explain in a few words the problem which will be under discussion, we remind that typical relativistic equations of
motion have singularity at some value of a particle speed. The singularity determines behavior of the particle in
ultra-relativistic limit.  For instance, the standard equations of spinless particle interacting with
electromagnetic field in the physical-time parametrization $x^\mu(t)=(ct, {\bf x}(t))$
\begin{eqnarray}\label{gm7.2}
\frac{d}{dt}\left(\frac{\dot x^\mu}{\sqrt{-\eta_{\mu\nu}\dot x^\mu\dot x^\nu}}\right)=\frac{e}{mc^2}F^\mu{}_\nu \dot
x^\nu,
\end{eqnarray}
become singular as the relativistic contraction factor vanishes, $\eta_{\mu\nu}\dot x^\mu\dot x^\nu=c^2-{\bf
v}^2=0$. Rewriting the equations in the form of second law of Newton we find an acceleration. For the case, the
longitudinal acceleration reads $a_{||}={\bf v} {\bf a}=\frac{e(c^2-{\bf v}^2)^{\frac{3}{2}}}{mc^3}({\bf E}{\bf v})$,
that is the factor, elevated in some degree, appears on the right hand side of the equation, and thus determines the
value of velocity at which the longitudinal acceleration vanishes, $a_{||}\stackrel{v\rightarrow c}{\longrightarrow}0$.
For the present case, the singularity implies that during its evolution in the external field, the spinless particle can not
exceed the speed of light $c$.

In the equations for spinning particle, instead of the original metric ($\eta_{\mu\nu}$ in flat and $g_{\mu\nu}$ in
curved space) emerges the effective metric $G_{\mu\nu}=g_{\mu\nu}+h_{\mu\nu}$, with spin and field-dependent
contribution $h_{\mu\nu}$. This turn out to be true both for MPTD and Frenkel equations. This leads to (drastic in some
cases) changes \cite{deriglazovMPL2015, DW2015.2} in behavior of spinning particle as compare with (\ref{gm7.2}). The
present work is devoted to detailed analysis of the behavior in ultra-relativistic regime.

We will use the following terminology. The speed $v_{cr}$ that a particle can not exceed during its evolution in an
external field is called critical speed\footnote{We prefer the term critical speed instead of maximum speed since
$v_{cr}$ generally is spin and field-dependent quantity, see below.}. The observer independent scale $c$ of special
relativity is called, as usual, the speed of light.

The work is organized as follows. In Sect. \ref{sec2} we define three-dimensional acceleration (\ref{La.8.5}) of a
particle in an arbitrary gravitational field. The definition guarantees that massive spinless particle propagating
along four-dimensional geodesic can not exceed the speed of light. Then we obtain the expressions (\ref{La.20.9}) and
(\ref{La.20.10}) for the acceleration implied by equation of a general form (\ref{La.20.8}). They will be repeatedly
used in the subsequent sections. In Sect. \ref{sec3} we shortly review the vector model of spin and present three
equivalent Lagrangians of the free theory. In Sect. \ref{sec4.1} we obtain equations of the particle minimally
interacting with gravity starting from the Lagrangian action without auxiliary variables. The variational problem leads
to the theory with fixed value of spin. In Sect. \ref{sec4.2} we present the Lagrangian which leads to the model of
Hanson-Regge type \cite{hanson1974}, with unfixed spin and with a mass-spin trajectory constraint. In Sect.
\ref{sec4.3} we present the MPTD-equations in the form convenient for our analysis, and show their equivalence with
those obtained in Sect. \ref{sec4.1}. In Sect. \ref{sec4.4} we discuss the problems arising in ultra-relativistic limit
of MPTD-equations. The first problem is the discrepancy between the critical speed with the speed of light.  We should
note that similar observations were mentioned in a number of works. The appearance of trajectories with space-like
four-velocity was remarked by Hanson and Regge in their model of spherical top in electromagnetic field
\cite{hanson1974}. Space-like trajectories of this model in gravitational fields were studied in \cite{Hojman,
Koch2015}. The second problem is that the transversal acceleration increases with velocity and blows up in the
ultra-relativistic limit.

In his work \cite{Khriplovich1989}, Khriplovich proposed non-minimal interaction of a rotating body through the
gravimagnetic moment $\kappa$. In Sect. \ref{sec5.1} we construct the non-minimal interaction starting from the
Hamiltonian variational problem, and show (Sect. \ref{sec5.2}) that the model with $\kappa=1$ has reasonable behavior
in ultra-relativistic limit. The Lagrangian with one auxiliary variable for the particle with gravimagnetic moment is
constructed in Sect. \ref{sec5.3}.  In Sect. \ref{sec6} we construct two toy models of spinless particle with critical
speed different from the speed of light. In sect. \ref{sec7.1} we analyze generalization of the Frenkel equations to
the case of a particle with magnetic moment in an arbitrary electromagnetic field in Minkowski space. Here we start
from the Lagrangian action with one auxiliary variable. In Sect. \ref{sec7.2} we show that critical speed of the
particle with anomalous magnetic moment is different from the speed of light, if we use the standard special-relativity
notions for time and distance. In Sect. \ref{sec7.3} we show that the equality between the two speeds can be preserved
assuming that three-dimensional geometry should be defined with respect to effective metric arised due to interaction
of spin with electromagnetic field. We point out that a possibility of deformed relation between proper and laboratory
time in the presence of electromagnetic field was discussed before by van Holten in his model of spin \cite{Holten}.

{\bf Notation.} Our variables are taken in arbitrary parametrization $\tau$, then $\dot x^\mu=\frac{dx^\mu}{d\tau}$.
Covariant derivative is $\nabla P^\mu=\frac{dP^\mu}{d\tau} +\Gamma^\mu_{\alpha\beta}\dot x^\alpha P^\beta$ and
curvature is $R^\sigma{}_{\lambda\mu\nu}=\partial_\mu\Gamma^\sigma{}_{\lambda\nu} -\partial_\nu
\Gamma^\sigma{}_{\lambda\mu}+\Gamma^\sigma{}_{\beta\mu}\Gamma^{\beta}{}_{\lambda\nu}-
\Gamma^\sigma{}_{\beta\nu}\Gamma^{\beta}{}_{\lambda\mu}$. The square brackets mean antisymmetrization,
$\omega^{[\mu}\pi^{\nu]}=\omega^\mu\pi^\nu-\omega^\nu\pi^\mu$. For the four-dimensional quantities we suppress the
contracted indexes and use the notation  $\dot x^\mu G_{\mu\nu}\dot x^\nu=\dot xG\dot x$,  $N^\mu{}_\nu\dot
x^\nu=(N\dot x)^\mu$, $\omega^2=g_{\mu\nu}\omega^\mu\omega^\nu$, $\mu, \nu=0, 1, 2, 3$.  Notation for the scalar
functions constructed from second-rank tensors are $\theta S= \theta^{\mu\nu}S_{\mu\nu}$, $S^2=S^{\mu\nu}S_{\mu\nu}$.

When we work in four-dimensional Minkowski space with coordinates $x^\mu=(x^0=ct,~  x^i)$, we use the metric
$\eta_{\mu\nu}=(-, +, +, +)$, then $\dot x\omega=\dot x^\mu\omega_\mu=-\dot x^0\omega^0+\dot x^i\omega^i$ and so on.
Suppressing the indexes of three-dimensional quantities, we use bold letters, $v^i\gamma_{ij}a^j={\bf v}\gamma{\bf a}$,
$v^iG_{i\mu}v^\mu={\bf v}Gv$, $i, j=1, 2, 3$, and so on.

Electromagnetic field:
\begin{eqnarray}\label{L.0}
F_{\mu\nu}=\partial_\mu A_\nu-\partial_\nu A_\mu=(F_{0i}=-E_i, ~ F_{ij}= \epsilon_{ijk}B_k), \cr
E_i=-\frac{1}{c}\partial_tA_i+\partial_i A_0, \quad B_i=\frac12\epsilon_{ijk}F_{jk}=\epsilon_{ijk}\partial_j A_k.
\end{eqnarray}

\section{Three-dimensional acceleration of spinless particle in general relativity}\label{sec2}

By construction of Lorentz transformations, the speed of light in special relativity is an observer-independent
quantity. As we have mentioned in the Introduction, the invariant scale is closely related with the critical speed in
an external field. In a curved space we need to be more careful since the three-dimensional geometry should respect the
coordinate independence of the speed of light. To achieve this, we use below the Landau and Lifshitz procedure
\cite{Landau:2} to define time interval, three-dimensional distance and velocity. Then we introduce the notion of
three-dimensional acceleration which guarantees that massive spinless particle propagating along four-dimensional
geodesic can not exceed the speed of light. The expression (\ref{La.20.10}) for longitudinal acceleration implied by
equation of the form (\ref{La.20.8}) will be repeatedly used in subsequent sections.

Consider an observer that labels the events by the coordinates $x^\mu$ of pseudo Riemann space \cite{Landau:2, Gour}
\begin{eqnarray}\label{La.1}
{\bf M}^{(1,3)}=\{x^\mu,  ~  g_{\mu\nu}(x^\rho),  ~  \mbox{sign} ~ g_{\mu\nu}=(-, +, +, +)\},
\end{eqnarray}
to describe the motion of a particle in gravitational field with metric $g_{\mu\nu}$. Formal definitions of
three-dimensional quantities subject to the discussion can be obtained representing interval in $1+3$ block-diagonal
form \cite{Landau:2}
\begin{eqnarray}\label{La.2}
-ds^2=g_{\mu\nu}dx^\mu dx^\nu= \qquad \qquad \qquad  \cr
-c^2\left[\frac{\sqrt{-g_{00}}}{c}(dx^0+\frac{g_{0i}}{g_{00}}dx^i)\right]^2+\left(g_{ij}-\frac{g_{0i}g_{0j}}{g_{00}}\right)dx^idx^j.
\end{eqnarray}
This prompts to introduce infinitesimal time interval, distance and speed as follows:
\begin{eqnarray}\label{La.3.0}
dt=\frac{\sqrt{-g_{00}}}{c}(dx^0+\frac{g_{0i}}{g_{00}}dx^i)\equiv-\frac{g_{0\mu}dx^\mu}{c\sqrt{-g_{00}}}.
\end{eqnarray}
\begin{eqnarray}\label{La.3}
dl^2=(g_{ij}-\frac{g_{0i}g_{0j}}{g_{00}})dx^idx^j\equiv\gamma_{ij}dx^idx^j, \qquad v=\frac{dl}{dt}.
\end{eqnarray}
Therefore the conversion factor between intervals of the world time $\frac{dx^0}{c}$ and the time $dt$ measured by
laboratory clock is

\begin{eqnarray}\label{La.3.1}
\frac{dt}{dx^0}=\frac{\sqrt{-g_{00}}}{c}(1+\frac{g_{0i}}{g_{00}}\frac{dx^i}{dx^0}).
\end{eqnarray}

Introduce also the three-velocity vector ${\bf v}$ with components
\begin{eqnarray}\label{La.5}
v^i=\left(\frac{dt}{dx^0}\right)^{-1}\frac{dx^i}{dx^0},
\end{eqnarray}
or, symbolically, $v^i=\frac{dx^i}{dt}$. We stress that contrary to $\frac{d}{dx^\mu}$, the set  $(\frac{d}{dt},
\frac{d}{dx^i})$ is non-holonomic basis of tangent space \footnote{Let $e_\mu=\tilde a^\alpha{}_\mu\partial_\alpha$ be
a basis of tangent space and $e^\mu=a^\mu{}_\alpha dx^\alpha$, where $a^\mu{}_\alpha\tilde
a^\alpha{}_\nu=\delta^\mu{}_\nu$, be the dual basis for $e_\mu$, that is $e^\mu(e_\nu)=\delta^\mu{}_\nu$. $e_\mu$ is
the holonomic basis (that is $e_\mu=\frac{\partial}{\partial x'^\mu}$ are tangent to some coordinate lines $x'^\mu$) if
$(e_\mu e_\nu-e_\nu e_\mu)f=0$. For the matrix $a^\mu{}_\alpha$ which determines the dual basis $e^\mu$, this condition
reduces to the simple equation $\partial_\mu a^\alpha{}_\nu-\partial_\nu a^\alpha{}_\mu=0$. For the matrix which
determines our $1+3$ decomposition we have $a^0{}_\mu=-\frac{g_{0\mu}}{c\sqrt{-g_{00}}}$, $a^i{}_0=0$, and
$a^i{}_j=\delta^i{}_j$, then, for instance, $\partial_\mu a^0{}_\nu-\partial_\nu
a^0{}_\mu=-\frac{1}{c\sqrt{-g_{00}}}(\partial_\mu g_{0\nu}-\partial_\nu g_{0\mu})\ne 0$. So the set
$(\frac{\partial}{\partial t}, \frac{\partial}{\partial x^i})$ generally does not represent a holonomic basis.}. This
does not represent any special problem for our discussion since we are interested in the differential quantities such
as velocity and acceleration.

The equation (\ref{La.5}) is consistent with the above definition of $v$: $v^2=\left(\frac{dl}{dt}\right)^2={\bf
v}^2=v^i\gamma_{ij}v^j$. In the result, the interval acquires the form similar to special relativity (but now we have
${\bf v}^2={\bf v}\gamma{\bf v}$)
\begin{eqnarray}\label{La.6}
-ds^2=-c^2dt^2+dl^2=-c^2dt^2\left(1-\frac{{\bf v}^2}{c^2}\right).
\end{eqnarray}
This equality holds in any coordinate system $x^\mu$. Hence a particle with the propagation law $ds^2=0$ has the speed
${\bf v}^2=c^2$, and this is a coordinate-independent statement.

For the latter use we also introduce the four-dimensional quantity
\begin{eqnarray}\label{La.5.1}
v^\mu=\left(\frac{dt}{dx^0}\right)^{-1}\frac{dx^\mu}{dx^0}=\left(\left(\frac{dt}{dx^0}\right)^{-1}, ~  {\bf v}\right).
\end{eqnarray}
Combining the equations (\ref{La.5}) and (\ref{La.3.1}), we can present the conversion factor in terms of
three-velocity as follows:
\begin{eqnarray}\label{La.5.2}
\left(\frac{dt}{dx^0}\right)^{-1}=v^0=\frac{c}{\sqrt{-g_{00}}}-\frac{g_{0i}v^i}{g_{00}}.
\end{eqnarray}

\begin{figure}[t] \centering
\includegraphics[width=200pt, height=100pt]{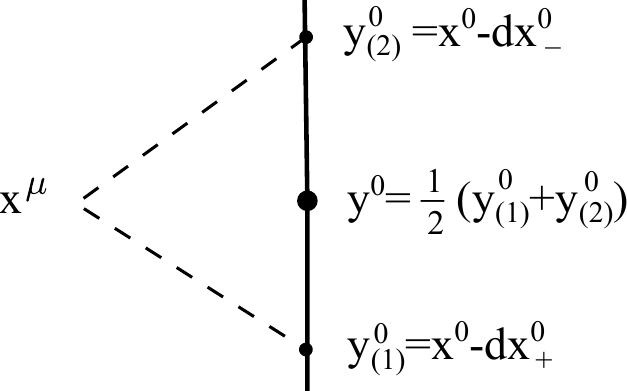}
\caption{Definition of simultaneous events. The vertical line represents a world-line of the laboratory clock. The
points $y^0_{(1)}$ and $y^0_{(2)}$ nave have null-interval with $x^{\mu}$. Then the middle point $y^0$ represents the
event simultaneous with $x^{\mu}$.}\label{ch06:fig6.7}
\end{figure}

\begin{figure}[t] \centering
\includegraphics[width=200pt, height=130pt]{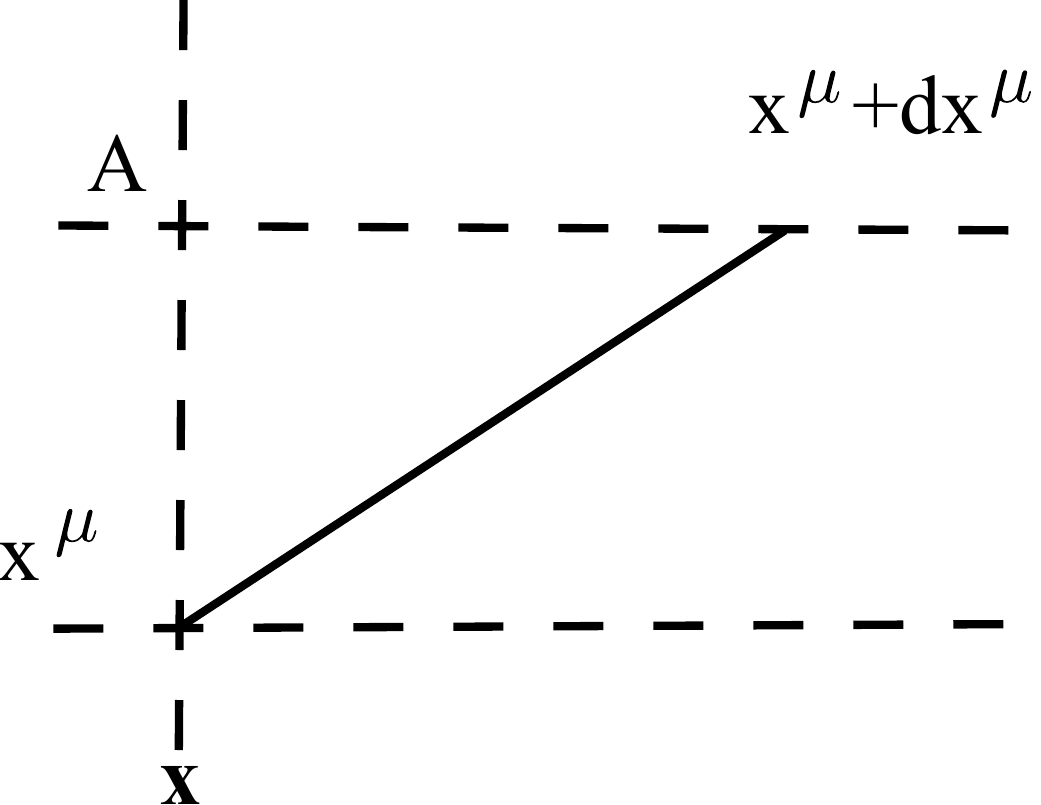}
\caption{Definition of simultaneous events. The vertical line represents a world-line of the laboratory clock. The
points $y^0_{(1)}$ and $y^0_{(2)}$ nave have null-interval with $x^{\mu}$. Then the middle point $y^0$ represents the
event simultaneous with $x^{\mu}$.}\label{ch06:fig6.8}
\end{figure}

These rather formal tricks are based \cite{Landau:2} on the notion of simultaneity in general relativity and on the
analysis of flat limit. Four-interval of special relativity has direct physical interpretation in two cases. First, for
two events which occur at the same point, the four-interval is proportional to time interval,
$dt=-\frac{ds}{c}$.
Second, for simultaneous events the four-interval coincides with distance,
$dl=ds$.
Assuming that the same holds in general relativity, let us analyze infinitesimal time interval and distance between two
events with coordinates $x^\mu$ and $x^\mu+dx^\mu$. The world line $y^\mu=(y^0, {\bf y}=\mbox{const})$ is associated
with laboratory clock placed at the spacial point ${\bf y}$. So the time-interval between the events $(y^0, {\bf y})$
and $(y^0+dy^0, {\bf y})$ measured by the clock is
\begin{eqnarray}\label{La.6.020}
dt=-\frac{ds}{c}=\frac{\sqrt{-g_{00}}}{c}dy^0.
\end{eqnarray}
Consider the event $x^\mu$ infinitesimally closed to the world line $(y^0,{\bf y}=\mbox{const})$. To find the event on
the world line which is simultaneous with $x^\mu$, we first look for the events $y^\mu_{(1)}$ and $y^\mu_{(2)}$ which
have null-interval with $x^\mu$, $ds(x^\mu, y^\mu_{(a)})=0$. The equation $g_{\mu\nu}dx^\mu dx^\nu=0$ with
$dx^\mu=x^\mu-y^\mu$ has two solutions $dx^0_{\pm}=\frac{g_{0i}dx^i}{-g_{00}}\pm\frac{\sqrt{d{\bf x}\gamma d{\bf
x}}}{\sqrt{-g_{00}}}$, then $y^0_{(1)}=x^0-dx^0_{+}$ and $y^0_{(2)}=x^0-dx^0_{-}$. Second, we compute the middle point
\begin{eqnarray}\label{La.6.1}
y^0=\frac{1}{2}(y^0_{(1)}+y^0_{(2)})=x^0+\frac{g_{0i}dx^i}{g_{00}}.
\end{eqnarray}
By definition\footnote{In the flat limit the sequence $y^\mu_{(1)}$, $x^\mu$, $y^\mu_{(2)}$ of events can be associated
with emission, reflection and absorbtion of a photon with the propagation law $ds=0$. Then the middle point
(\ref{La.6.1}) should be considered simultaneous with $x^0$.}, the event $(y^0, {\bf y})$ with the null-coordinate
(\ref{La.6.1}) is simultaneous with the event $(x^0, {\bf x})$, see Figure \ref{ch06:fig6.7}.
By this way we synchronized clocks at the spacial points ${\bf x}$ and ${\bf y}$. According to (\ref{La.6.1}), the
simultaneous events have different null-coordinates, and the difference $dx^0$ obeys the equation
\begin{eqnarray}\label{La.6.2}
dx^0+\frac{g_{0i}dx^i}{g_{00}}=0.
\end{eqnarray}
Consider a particle which propagated from $x^\mu$ to $x^\mu+dx^\mu$. Let us compute time-interval and distance between
these two events. According to (\ref{La.6.1}),  the event
\begin{eqnarray}\label{La.6.3}
\left(x^0+dx^0+\frac{g_{0i}dx^i}{g_{00}}, ~  {\bf x}\right),
\end{eqnarray}
at the spacial point ${\bf x}$ is simultaneous with $x^\mu+dx^\mu$, see Figure \ref{ch06:fig6.8}.
According to (\ref{La.6.020}) and (\ref{La.6.1}), the time interval between the events $x^\mu$ and (\ref{La.6.3}) is
\begin{eqnarray}\label{La.6.4}
dt=\frac{\sqrt{-g_{00}}}{c}(dx^0+\frac{g_{0i}}{g_{00}}dx^i).
\end{eqnarray}
Since the events $x^\mu+dx^\mu$ and (\ref{La.6.3}) are simultaneous, this equation gives also the time interval between
$x^\mu$ and  $x^\mu+dx^\mu$. Further, the difference of coordinates between the events $x^\mu+dx^\mu$ and
(\ref{La.6.3}) is $dz^\mu=(-\frac{g_{0i}dx^i}{g_{00}}, dx^i)$. As they are simultaneous, the distance between them is
\begin{eqnarray}\label{La.6.5}
dl^2=-ds^2=g_{\mu\nu}dz^\mu dz^\nu=(g_{ij}-\frac{g_{0i}g_{0j}}{g_{00}})dx^idx^j\equiv\gamma_{ij}dx^idx^j.
\end{eqnarray}
Since (\ref{La.6.3}) occur at the same spacial point as $x^\mu$, this equation gives also the distance between $x^\mu$
and $x^\mu+dx^\mu$. The equations (\ref{La.6.4}) and (\ref{La.6.5}) coincide with the formal definitions presented above, Eqs.
(\ref{La.3.0}) and (\ref{La.3}).

We now turn to the definition of three-acceleration. The spinless particle in general relativity follows a geodesic
line. If we take the proper time to be the parameter, geodesics obey the system
\begin{eqnarray}\label{L.6.10}
\nabla_s\frac{dx^\mu}{ds}\equiv\frac{d^2x^\mu}{ds^2}+\Gamma^\mu{}_{\alpha\beta}\frac{dx^\alpha}{ds}\frac{dx^\beta}{ds}=0,
\qquad g_{\mu\nu}\frac{dx^\mu}{ds}\frac{dx^\nu}{ds}=-1,
\end{eqnarray}
where
\begin{eqnarray}\label{L.6.10.1}
\Gamma^\mu{}_{\alpha\beta}=\frac12g^{\mu\nu}(\partial_\alpha g_{\nu\beta}+\partial_\beta g_{\alpha\nu}-\partial_\nu g_{\alpha\beta}).
\end{eqnarray}
Due to this definition, the system (\ref{L.6.10}) obeys the identity
$g_{\mu\nu}\frac{dx^\mu}{ds}\nabla_s\frac{dx^\nu}{ds}=0$.

The system in this parametrization has no sense for the case we are interested in, $ds^2\rightarrow 0$. So we rewrite it in arbitrary
parametrization $\tau$
\begin{eqnarray}\label{L.6.11}
\frac{d\tau}{ds}\frac{d}{d\tau}\left(\frac{d\tau}{ds}\frac{d
x^\mu}{d\tau}\right)+\left(\frac{d\tau}{ds}\right)^2
\Gamma^\mu{}_{\alpha\beta}(g)\frac{dx^\alpha}{d\tau}\frac{dx^\beta}{d\tau}=0, \qquad
\frac{d\tau}{ds}=\frac{1}{\sqrt{-\dot xg\dot x}}, \qquad \qquad \qquad
\end{eqnarray}
this yields the equation of geodesic line in reparametrization-invariant form
\begin{eqnarray}\label{L.6.12}
\frac{1}{\sqrt{-\dot xg\dot x}}\frac{d}{d\tau}\left(\frac{\dot x^\mu}{\sqrt{-\dot xg\dot
x}}\right)=-\Gamma^\mu{}_{\alpha\beta}(g) \frac{\dot x^\alpha}{\sqrt{-\dot xg\dot x}}\frac{\dot x^\beta}{\sqrt{-\dot
xg\dot x}}.
\end{eqnarray}

The formalism (\ref{La.3.0})-(\ref{La.6}) remains manifestly
covariant under subgroup of spacial transformations $x^0=x'{}^0$, $x^i=x^i(x'{}^j)$, $\frac{\partial x^i}{\partial
x'{}^j}\equiv a^i{}_j(x')$. Under these transformations $g_{00}$ is a scalar function, $g_{0i}$ is a vector while
$g_{ij}$ and $\gamma_{ij}$ are tensors. Since $g^{ij}\gamma_{jk}=\delta^{i}{}_k$, the inverse metric of  $\gamma_{ij}$
turns out to be $(\gamma^{-1})^{ij}=g^{ij}$. Introduce the covariant derivatives $\nabla_k$ of a vector field
$\xi^i(x^0, x^k)$
\begin{eqnarray}\label{La.8}
\nabla_k\xi^i=\partial_k\xi^i+\tilde\Gamma^i{}_{kj}(\gamma)\xi^j.
\end{eqnarray}
The three-dimensional Christoffel symbols $\tilde\Gamma^i{}_{jk}(\gamma)$ are constructed with help of three-dimensional metric
$\gamma_{ij}(x^0, x^k)$ written in Eq. (\ref{La.3}), where $x^0$ is considered as a parameter
\begin{eqnarray}\label{La.8.1}
\tilde\Gamma^i{}_{jk}(\gamma)=\frac12\gamma^{ia}(\partial_j\gamma_{ak}+\partial_k\gamma_{aj}-\partial_a\gamma_{jk}).
\end{eqnarray}
As a consequence, the metric $\gamma$ is covariantly constant, $\nabla_k\gamma_{ij}=0$.

The velocity (\ref{La.5}) behaves as a vector,
$v^i(x^0)=a^i{}_j(x'{}^k(x^0))v'{}^j(x^0)$,
so below we use also the covariant derivative
\begin{eqnarray}\label{La.8}
\nabla_0v^i=\frac{dv^i}{dx^0}+\tilde\Gamma^i{}_{jk}(\gamma)\frac{dx^j}{dx^0}v^k.
\end{eqnarray}

We associated with ${\bf M}^{(1,3)}$ the one-parameter family of three-dimensional spaces ${\bf M}^3_{x^0}=\{x^k, ~
\gamma_{ij}, ~  \nabla_k\gamma_{ij}=0\}$. Note that velocity has been defined above as a tangent vector to the curve
which cross the family and is parameterized by this parameter, $x^i(x^0)$.

To define an acceleration of a particle in the three-dimensional geometry, we need the notion of a constant vector
field (or, equivalently, the parallel-transport equation). In the case of stationary field, $g_{\mu\nu}(x^k)$, we can
identify the curve $x^i(x^0)$ of ${\bf M}^{(1,3)}$ with that of any one of ${\bf M}^3_{x^0}=\{x^k, ~
\gamma_{ij}(x^k)\}$.  So we have the usual three-dimensional Riemann geometry, and an analog of constant vector field
of Euclidean geometry is the covariantly-constant field along the line $x^i(x^0)$, $\nabla_0\xi^i=0$. For the field of
velocity, its deviation from the covariant constancy is the acceleration \cite{deriglazovMPL2015}

\begin{eqnarray}\label{La.8.2}
a^i=
\left(\frac{dt}{dx^0}\right)^{-1}\nabla_0v^i=\left(\frac{dt}{dx^0}\right)^{-1}\frac{dv^i}{dx^0}+\tilde\Gamma^i{}_{jk}v^jv^k.
\end{eqnarray}

To define an acceleration in general case, $\gamma_{ij}(x^0, x^i)$, we need to adopt some notion of a constant vector
field along the trajectory $x^i(x^0)$ that cross the family ${\bf M}^3_{x^0}$. We propose the definition which preserves one
of basic properties of constant fields in differential geometry. In Euclidean and Minkowski spaces, the canonical scalar product of
two constant fields does not depend on the point where it was computed. In (pseudo) Riemann space, constant vector field is defined in such a way that the same property holds \cite{deriglazov2010classical}. In particular, taking the scalar product along
a line $x^i(x^0)$, we have $\frac{d}{dx^0}(\xi, \eta)=0$. For the constant fields in the three-dimensional geometry resulting after Landau-Lifshitz $1+3$\,-decomposition,  we demand
the same (necessary) condition: $\frac{d}{dx^0}[\xi^i(x^0)\gamma_{ij}(x^0, x^i(x^0))\eta^i(x^0)]=0$. Taking into
account that $\nabla_k\gamma_{ij}=0$, this condition can be written as follows
\begin{eqnarray}\label{La.8.3}
(\nabla_0\xi+\frac12\xi\partial_0\gamma\gamma^{-1}, \eta)+(\xi, \nabla_0\eta+\frac12\gamma^{-1}\partial_0\gamma\eta)=0.
\nonumber
\end{eqnarray}
This equation is satisfied, if we take the parallel-transport equation to be
\begin{eqnarray}\label{La.8.4}
\nabla_0\xi^i+\frac12(\xi\partial_0\gamma\gamma^{-1})^i=0.
\end{eqnarray}
Deviation from the constant field is an acceleration. So we define acceleration with respect to physical time as
follows:
\begin{eqnarray}\label{La.8.5}
a^i=\left(\frac{dt}{dx^0}\right)^{-1}\left[\nabla_0v^i+\frac12({\bf v}\partial_0\gamma\gamma^{-1})^i\right].
\end{eqnarray}
For the special case of stationary field, $g_{\mu\nu}(x^i)$, the definition (\ref{La.8.5}) reduces to (\ref{La.8.2}) and to that
of Landau and Lifshitz, see page 251 in \cite{Landau:2}.

The extra-term appeared in this equation plays an essential role to provide that for the geodesic motion we have:
$a_{||}{\stackrel{v\rightarrow c}\longrightarrow} 0$. As a consequence, geodesic particle in gravitational field can not
exceed the speed of light. To show this, we compute the longitudinal acceleration $({\bf v}\gamma{\bf a})$ implied by
geodesic equation (\ref{L.6.12}). Take $\tau=x^0$, then $\sqrt{-\dot xg\dot x}=\frac{dt}{dx^0}\sqrt{c^2-{\bf
v}\gamma{\bf v}}$, and spacial part of (\ref{L.6.12}) is
\begin{eqnarray}\label{La.12}
\left(\frac{dt}{dx^0}\right)^{-1}\frac{d}{dx^0}\frac{v^i}{\sqrt{c^2-{\bf v}\gamma{\bf
v}}}=\frac{f^i}{\sqrt{c^2-{\bf v}\gamma{\bf v}}},
\end{eqnarray}
where
\begin{eqnarray}\label{La.13}
f^i(v^\mu)=-\left(\frac{dt}{dx^0}\right)^{-2}\Gamma^i{}_{00}-\Gamma^i{}_{jk}v^jv^k-
2\left(\frac{dt}{dx^0}\right)^{-1}\Gamma^i{}_{0k}v^k=-\Gamma^i_{\mu\nu}v^\mu v^\nu, \qquad
\end{eqnarray}
is non-singular function as $v\rightarrow c$. Computing derivative on the l.h.s. of  (\ref{La.12}), we complete
$\frac{dv^i}{dx^0}$ up to covariant derivative $\nabla_0 v^i$
\begin{eqnarray}\label{La.13.1}
\frac{d}{dx^0}\frac{v^i}{\sqrt{c^2-{\bf v}\gamma{\bf v}}}=\nabla_0
v^i-\tilde\Gamma^i{}_{jk}(\gamma)v^jv^k\frac{dt}{dx^0}+\frac{v^i}{2(c^2-{\bf v}\gamma{\bf v})}\frac{d}{dx^0}({\bf
v}\gamma{\bf v}).
\end{eqnarray}
For the derivative contained in the last term we find, using covariant constancy of $\gamma$
\begin{eqnarray}\label{La.13.2}
\frac{d}{dx^0}[{\bf v}\gamma(x^0, x^i){\bf v}]=2{\bf v}\gamma\nabla_0{\bf v}+{\bf v}\partial_0\gamma{\bf v}+{\bf
v}\nabla_0\gamma{\bf v}=2{\bf v}\gamma\nabla_0{\bf v}+{\bf v}\partial_0\gamma{\bf v}.
\end{eqnarray}
Then (\ref{La.12}) acquires the form
\begin{eqnarray}\label{La.14}
\left(\frac{dt}{dx^0}\right)^{-1}\left[M^i{}_j\nabla_0v^j+\frac{({\bf v}\partial_0\gamma{\bf v})}{2(c^2-{\bf v}\gamma
{\bf v})}v^i\right]=f^i+\tilde\Gamma^i{}_{kl}v^kv^l,
\end{eqnarray}
where
\begin{eqnarray}\label{La.17}
M^i{}_j=\delta^i{}_j+\frac{v^i({\bf v}\gamma)_j}{c^2-{\bf v}\gamma{\bf v}}.
\end{eqnarray}
We apply the inverse matrix
\begin{eqnarray}\label{La.17.1}
\tilde M^i{}_j=\delta^i{}_j-\frac{v^i({\bf v}\gamma)_j}{c^2},
\end{eqnarray}
and use the identity
\begin{eqnarray}\label{La.17.2}
\tilde M^i{}_jv^j=\frac{c^2-{\bf v}\gamma{\bf v}}{c^2}v^i,
\end{eqnarray}
then
\begin{eqnarray}\label{La.14.1}
\left(\frac{dt}{dx^0}\right)^{-1}\left[\nabla_0v^i+\frac{({\bf v}\partial_0\gamma{\bf v})}{2c^2}v^i\right]=\tilde
M^i{}_j\left[f^j+\tilde\Gamma^j{}_{kl}v^kv^l\right].
\end{eqnarray}
Next,  we complete $\nabla_0v^i$ up to the acceleration (\ref{La.8.5}). Then (\ref{La.14.1}) yields
\begin{eqnarray}\label{La.11}
a^i=\frac12\left(\frac{dt}{dx^0}\right)^{-1}\left[({\bf v}\partial_0\gamma\gamma^{-1})^i-\frac{({\bf
v}\partial_0\gamma{\bf v})}{c^2}v^i\right]+\tilde M^i{}_j[-\Gamma^j_{\mu\nu}v^\mu v^\nu+\tilde\Gamma^j{}_{kl}(\gamma)v^kv^l].
\end{eqnarray}
Contracting this with $({\bf v}\gamma)_i$, we use $({\bf v}\gamma)_i\tilde M^i{}_j=\frac{c^2-{\bf v}\gamma{\bf
v}}{c^2}({\bf v}\gamma)_j$ and obtain longitudinal acceleration
\begin{eqnarray}\label{La.20}
{\bf v}\gamma{\bf a}=\frac12\left(\frac{dt}{dx^0}\right)^{-1}\left[({\bf v}\partial_0\gamma{\bf v})-({\bf
v}\partial_0\gamma{\bf v})\frac{({\bf v}\gamma{\bf v})}{c^2}\right]+\left(1-\frac{{\bf v}\gamma{\bf
v}}{c^2}\right)({\bf v}\gamma)_i[-\Gamma^i_{\mu\nu}v^\mu v^\nu+\tilde\Gamma^i{}_{kl}(\gamma)v^kv^l].
\end{eqnarray}
This implies ${\bf v}\gamma{\bf a}\rightarrow 0$ as ${\bf v}\gamma{\bf v}\rightarrow c^2$.

The last term in (\ref{La.8.5}) yields the important factor $({\bf v}\partial_0\gamma{\bf v})$ in Eq. (\ref{La.20}). As
the equations of motion  (\ref{La.11}) and (\ref{La.20}) do not contain the square root $\sqrt{c^2-{\bf v}\gamma{\bf
v}}$, they have sense even for $v>c$. Without this factor, we would have ${\bf v}\gamma{\bf a}\ne 0$ as ${\bf
v}\gamma{\bf v}\rightarrow c^2$, so the particle in gravitational field could exceed $c$ and then continues accelerate.
The same happen if we try to define acceleration using usual derivative instead of the covariant one\footnote{Indeed, instead of Eq. (\ref{La.8.5}), let us define an acceleration according to the expression $a'^i=\left(\frac{dt}{dx^0}\right)^{-1}\left[\frac{dv^i}{dx^0}+\frac12({\bf v}\partial_0\gamma\gamma^{-1})^i\right]$. Then for the geodesic particle we obtain, instead of (\ref{La.20}), the longitudinal acceleration ${\bf v}\gamma{\bf a}'=\mbox{[r.h.s. of (\ref{La.20})]}-({\bf v}\gamma)_i \tilde\Gamma^i{}_{jk}(\gamma)\frac{dx^j}{dx^0}v^k=\mbox{[r.h.s. of (\ref{La.20})]}-\frac{1}{2}\partial_i\gamma_{jk}v^iv^jv^k$. The extra term does not involve the factor $c^2-{\bf v}\gamma{\bf v}$ and so does not vanish at $|v|=c$.}.

Let us confirm that $c$ is the only special point of the function (\ref{La.20}) representing the longitudinal acceleration.
Using Eqs. (\ref{L.6.10.1}), (\ref{La.3})-(\ref{La.5.1}), (\ref{La.8.1}) and the identities
\begin{eqnarray}\label{La.20.1}
\gamma_{ij}g^{jk}=\delta_i{}^k, \qquad \gamma_{ij}g^{j0}=-\frac{g_{0i}}{g_{00}},
\end{eqnarray}
we can present the right hand side of Eq. (\ref{La.20}) in terms of initial metric as follows
\begin{eqnarray}\label{La.20.2}
{\bf v}\gamma{\bf a}=\frac{c^2-{\bf v}\gamma{\bf
v}}{2c\sqrt{-g_{00}}}\left\{\frac{c}{\sqrt{-g_{00}}}[\left(\frac{dt}{dx^0}\right)^{-1}\partial_0
g_{00}+v^k\partial_kg_{00}]-\partial_0 g_{00}\left(\frac{dt}{dx^0}\right)^{-2}-2\partial_0
g_{0k}\left(\frac{dt}{dx^0}\right)^{-1}v^k-\partial_0 g_{kl}v^kv^l\right\}\equiv \cr \frac{c^2-{\bf v}\gamma{\bf
v}}{2c\sqrt{-g_{00}}}\left\{\frac{c}{\sqrt{-g_{00}}}v^\mu\partial_\mu g_{00}-\partial_0g_{\mu\nu}v^\mu v^\nu\right\}.
\qquad \qquad \qquad \qquad \qquad \qquad
\end{eqnarray}
The quantity $v^\mu$ has been defined in (\ref{La.5.1}). Excluding $v^0$ according to this expression, we obtain
\begin{eqnarray}\label{La.20.3}
{\bf v}\gamma{\bf a}=\frac{c^2-{\bf v}\gamma{\bf v}}{2\sqrt{-g_{00}}}\left\{\frac{v^k\partial_k
g_{00}}{\sqrt{-g_{00}}}-2\partial_0\left(\frac{g_{0i}}{\sqrt{-g_{00}}}\right)v^i-\frac{1}{c}\partial_0\gamma_{ij}v^i
v^j\right\}.
\end{eqnarray}
For the stationary metric, $g_{\mu\nu}(x^k)$, the equation (\ref{La.20.3}) acquires a specially simple form
\begin{eqnarray}\label{La.20.4}
{\bf v}\gamma{\bf a}=-(c^2-{\bf v}\gamma{\bf v})\frac{v^k\partial_kg_{00}}{2g_{00}}.
\end{eqnarray}
This shows that the longitudinal acceleration has only one special point in the stationary gravitational field, ${\bf v}\gamma{\bf a}\rightarrow 0$ as ${\bf v}\gamma{\bf v}\rightarrow c^2$. Then the same is true in general case (\ref{La.20.2}), at least for the metric which is sufficiently slowly varied in time.


While we have discussed the geodesic equation, the computation which leads to the formula (\ref{La.20}) can be repeated
for a more general equation. Let us formulate the result which will be repeatedly used below.
Using the factor $\sqrt{-\dot xg\dot x}$ we construct the reparametrization-invariant derivative
\begin{eqnarray}\label{La.20.7}
D=\frac{1}{\sqrt{-\dot xg\dot x}}\frac{d}{d\tau}.
\end{eqnarray}
Consider the reparametrization-invariant equation of the form
\begin{eqnarray}\label{La.20.8}
DDx^\mu(\tau)={\mathcal F}^\mu(Dx^\nu, \ldots ).
\end{eqnarray}
and suppose that the three-dimensional geometry is defined by $g_{\mu\nu}$. Then Eq. (\ref{La.20.8})
implies the three-acceleration
\begin{eqnarray}\label{La.20.9}
a^i=\tilde M^i{}_j\left[(c^2-{\bf v}\gamma{\bf v}){\mathcal F}^j+
\tilde\Gamma^j{}_{kl}(\gamma)v^kv^l\right]+\frac12\left(\frac{dt}{dx^0}\right)^{-1}\left[({\bf
v}\partial_0\gamma\gamma^{-1})^i-\frac{v^i}{c^2}({\bf v}\partial_0\gamma{\bf v})\right],
\end{eqnarray}
and the longitudinal acceleration
\begin{eqnarray}\label{La.20.10}
{\bf v}\gamma{\bf a}=\frac{(c^2-{\bf v}\gamma{\bf v})^2}{c^2}({\bf v}\gamma{\boldsymbol{\mathcal F}})+\frac{c^2-{\bf v}\gamma{\bf v}}{c^2}
\left[({\bf v}\gamma)_i\tilde\Gamma^i{}_{kl}(\gamma)v^kv^l+\frac12\left(\frac{dt}{dx^0}\right)^{-1}({\bf v}\partial_0\gamma{\bf v})\right].
\end{eqnarray}
The spacial part of the force is ${\mathcal F}^i={\mathcal F}^i(\frac{v^\nu}{\sqrt{c^2-{\bf v}\gamma{\bf v}}})$, where $v^\mu$
is given by (\ref{La.5.1}), and the connection $\tilde\Gamma^i{}_{kl}(\gamma)$ is constructed with help of the
three-dimensional metric $\gamma_{ij}=(g_{ij}-\frac{g_{0i}g_{0j}}{g_{00}})$ according to (\ref{La.8.1}). For the
geodesic equation in this notation we have ${\mathcal F}^i=-\Gamma^i{}_{\mu\nu}\frac{v^\mu v^\nu}{c^2-{\bf v}\gamma{\bf
v}}$. With this ${\mathcal F}^i$ the equations (\ref{La.20.9}) and (\ref{La.20.10}) coincide with (\ref{La.11}) and
(\ref{La.20}).

\section{Vector model of relativistic spin}\label{sec3}

The variational problem for vector model of spin interacting with electromagnetic and
gravitational fields can be formulated with various sets of auxiliary variables \cite{DPW2, DPM2, DPM3,
deriglazov2014Monster, DW2015.1}. For the free theory in flat space there is the Lagrangian action without auxiliary
variables. Configuration space consist of the position $x^\mu(\tau)$ and non Grassmann vector $\omega^\mu(\tau)$
attached to the point $x^\mu$. The action reads \cite{deriglazov2014Monster, DW2015.2}
\begin{eqnarray}\label{Lfree}
S=-\frac{1}{\sqrt{2}} \int d\tau \sqrt{m^2c^2 -\frac{\alpha}{\omega^2}} ~  \sqrt{-\dot x N \dot x - \dot\omega N
\dot\omega +\sqrt{[\dot x N\dot x + \dot\omega N \dot\omega]^2- 4 (\dot x N \dot\omega )^2}}.
\end{eqnarray}
The matrix $N_{\mu\nu}$ is the projector on the plane orthogonal to $\omega^\nu$
\begin{equation}\label{FF0}
N_{\mu\nu}= \eta_{\mu\nu}-\frac{\omega_\mu \omega_\nu}{\omega^2}, \quad \mbox{then} \quad
N_{\mu\alpha}N^{\alpha\nu}=N_\mu{}^\nu, \quad N_{\mu\nu} \omega^\nu=0.
\end{equation}
The double square-root structure in the expression (\ref{Lfree}) seem to be typical for the vector models of spin
\cite{hanson1974, Star2008}. This yields the primary constraint $T_4$ in (\ref{primary}) and,  at the end, the
supplementary spin condition (\ref{condition1}). The parameter $m$ is mass, while $\alpha$ determines the value of
spin. The value $\alpha=\frac{3\hbar^2}{4}$ corresponds to an elementary spin one-half particle. The model is
invariant\footnote{The reparametrizations are $\tau\rightarrow\tau'(\tau)$, $x'^\mu(\tau')=x^\mu(\tau)$,
$\omega'^\mu(\tau')=\omega^\mu(\tau)$, that is both $x$ and $\omega$ are scalar functions. The local spin-plane
transformations act in the plane determined by the vectors $\omega^\mu$ and $\pi^\nu$, see \cite{DPM1}.} under
reparametrizations and local spin-plane symmetries \cite{DPM1}.

The spin is described by Frenkel spin-tensor \cite{Frenkel}. In our model this is a composite quantity constructed from
$\omega^\mu$ and its conjugated momentum $\pi_\mu=\frac{\partial L}{\partial\dot\omega^\mu}$ as follows:
\begin{eqnarray}\label{FF.2}
S^{\mu\nu}=2(\omega^\mu\pi^\nu-\omega^\nu\pi^\mu)=(S^{i0}=D^i, ~ S_{ij}=2\epsilon_{ijk}S_k),
\end{eqnarray}
then $S_i=\epsilon_{ijk}\omega_j\pi_k=\frac{1}{4}\epsilon_{ijk}S_{jk}$. Here $S_i$ is three-dimensional spin-vector and
$D_i$ is dipole electric moment \cite{ba1}. In contrast to its constituents $\omega^\mu$ and $\pi^\nu$, the spin-tensor
is invariant under local spin-plane symmetry, and thus represents an observable quantity. Canonical quantization of the
model yields one-particle sector of the Dirac equation \cite{DPM2}.

In the formulation (\ref{Lfree}), the model admits minimal interaction with electromagnetic field and with gravity.
This does not spoil the number and the algebraic structure of constraints presented in the free theory. To describe the
spinning particle with magnetic and gravimagnetic moments, we will need the following two reformulations.

In the spinless limit, $\alpha=0$ and $\omega^\mu=0$, the functional (\ref{Lfree}) reduces to the standard expression,
$-mc\sqrt{-\dot x^\mu\dot x_\mu}$. The latter can be written in equivalent form using the auxiliary variable
$\lambda(\tau)$ as follows: $\frac{1}{2\lambda}\dot x^2-\frac{\lambda}{2}m^2c^2$. Similarly to this, (\ref{Lfree}) can
be presented in the equivalent form
\begin{eqnarray}\label{FF.1}
L=\frac{1}{4\lambda_1}\left[\dot xN\dot x+\dot\omega N\dot\omega-\sqrt{[\dot x N\dot x + \dot\omega N \dot\omega]^2- 4
(\dot x N \dot\omega )^2}\right]-\frac{\lambda_1}{2}[(mc)^2-\frac{\alpha}{\omega^2}].
\end{eqnarray}
In this formulation our model admits interaction of spin  with an arbitrary electromagnetic field through the magnetic
moment, see section \ref{sec7.1}.
Another form of the Lagrangian is
\begin{eqnarray}
L=-\sqrt{(mc)^2 -\frac{\alpha}{\omega^2}} ~ \sqrt{(1-\lambda^2)^{-1}\left[-\dot x N \dot x - \dot\omega N
\dot\omega + 2\lambda\dot x N \dot\omega \right]} \equiv \label{FF.1.1} \\
-\sqrt{(mc)^2 -\frac{\alpha}{\omega^2}} ~ \sqrt{-(N\dot x, N\dot\omega)\left(
\begin{array}{cc}
\frac{\eta}{1-\lambda^2}& \frac{-\lambda\eta}{1-\lambda^2}\\
\frac{-\lambda\eta}{1-\lambda^2}& \frac{\eta}{1-\lambda^2}
\end{array}
\right)\left(
\begin{array}{c}
N\dot x\\
N\dot\omega
\end{array}
\right)} \,. \label{FF.1.2}
\end{eqnarray}
Its advantage is that the expression under the square root represents quadratic form with respect the velocities $\dot
x$ and $\dot\omega$. To relate the  Lagrangians  (\ref{Lfree}) and (\ref{FF.1.1}), we exclude $\lambda$ from the
latter. Computing variation of (\ref{FF.1.1}) with respect to $\lambda$, we obtain the equation
\begin{eqnarray}\label{FF.1.3}
(\dot x N\dot\omega)\lambda^2-(\dot x N\dot x+\dot\omega N\dot\omega)\lambda+(\dot x N\dot\omega)=0,
\end{eqnarray}
which determines $\lambda$
\begin{eqnarray}\label{FF.1.4}
\lambda_{\pm}=\frac{(\dot x N\dot x+\dot\omega N\dot\omega)\pm \sqrt{(\dot x N\dot x+\dot\omega N\dot\omega)^2-4(\dot x
N\dot\omega)^2}} {2(\dot x N\dot\omega)}.
\end{eqnarray}
We substitute $\lambda_{+}$ into (\ref{FF.1.1}) and use $\lambda_{+}\lambda_{-}=1$, then (\ref{FF.1.1}) turns into
(\ref{Lfree}). In the formulation (\ref{FF.1.1}) our model admits interaction of spin  with gravity through the
gravimagnetic moment, see section \ref{sec5.3}.

\section{Minimal interaction with an arbitrary gravitational field}\label{sec4}

\subsection{Lagrangian and Hamiltonian formulations}\label{sec4.1}

The minimal interaction with gravitational field can be achieved by covariantization of the formulation without
auxiliary variables. In the expressions (\ref{Lfree}) and (\ref{FF0}) we replace $\eta_{\mu\nu}\rightarrow g_{\mu\nu}$,
and usual derivative by the covariant one, $\dot\omega^\mu\rightarrow \nabla\omega^\mu=\frac{d\omega^\mu}{d\tau}
+\Gamma^\mu_{\alpha\beta}\dot x^\alpha\omega^\beta$. Thus our Lagrangian in a curved background reads \cite{DW2015.1}
\begin{eqnarray}\label{L-curved}
L =-\frac{1}{\sqrt{2}} \left[ m^2c^2 -\frac{\alpha}{\omega^2} \right]^{\frac 12} ~  \sqrt{-\dot x N \dot x -
\nabla\omega N \nabla\omega +\sqrt{[\dot x N\dot x + \nabla\omega N \nabla\omega]^2- 4 (\dot x N \nabla\omega )^2}} \cr
\equiv -\frac{1}{\sqrt{2}} \left[ m^2c^2 -\frac{\alpha}{\omega^2} \right]^{\frac 12} ~ L_0. \qquad \qquad \qquad \qquad
\qquad \qquad \qquad \qquad \qquad  \qquad \qquad \qquad
\end{eqnarray}
Velocities $\dot x^\mu$, $\nabla\omega^\mu$ and projector $N_{\mu\nu}$ transform like contravariant vectors and
covariant tensor, so the action is manifestly invariant under general-coordinate transformations.

Let us construct Hamiltonian formulation of the model (\ref{L-curved}). Conjugate momenta for $x^\mu$ and $\omega^\mu$
are $p_\mu=\frac{\partial L}{\partial\dot x^\mu}$ and  $\pi_\mu=\frac{\partial L}{\partial\dot\omega^\mu}$
respectively. Due to the presence of Christoffel symbols in $\nabla\omega^\mu$, the conjugated momentum $p_\mu$ does
not transform as a vector, so it is convenient to introduce the canonical momentum

\begin{equation}
P_\mu\equiv p_\mu-\Gamma^\beta_{\alpha\mu}\omega^\alpha\pi_\beta \, ,
\end{equation}
the latter  transforms as a vector under general transformations of coordinates. Manifest form of the momenta is as
follows:
\begin{eqnarray}\label{p1}
P_\mu& =& \frac{1}{\sqrt{2}L_0}\left[m^2c^2 -\frac{\alpha}{\omega^2}\right]^{\frac12}\left[ N_{\mu\nu} \dot x^\nu -
K_\mu \right] ,
\end{eqnarray}
\begin{eqnarray}\label{p2}
\pi_\mu &=& \frac{1}{\sqrt{2}L_0}\left[m^2c^2 -\frac{\alpha}{\omega^2}\right]^{\frac12} \left[ N_{\mu\nu}
\nabla\omega^\nu - R_\mu \right]  ,
\end{eqnarray}
with
\begin{eqnarray}
K_\mu=T^{-1/2} \left[ (\dot x N \dot x +  \nabla\omega N \nabla\omega) (N \dot x)_\mu -
2 (\dot x N \nabla\omega) (N \nabla\omega)_\mu \right] \, , \nonumber \\
R_\mu=T^{-1/2} \left[ (\dot x N \dot x +  \nabla\omega N \nabla\omega) (N \nabla\omega)_\mu - 2 (\dot x N \nabla\omega)
(N \dot x)_\mu \right] \, . \nonumber
\end{eqnarray}
These vectors  obey the following remarkable  identities
\begin{eqnarray}\label{rel-1}
K^2= \dot x N\dot x \, , \quad R^2=\nabla \omega N \nabla \omega \, , \quad KR = -\dot xN\nabla \omega \, , \quad \dot
x R +\nabla \omega K =0 , \cr K\dot x  + R \nabla \omega =\sqrt{[\dot x N\dot x + \nabla\omega N \nabla\omega]^2- 4
(\dot x N \nabla\omega )^2} \, . \qquad \qquad
\end{eqnarray}
Using  (\ref{FF0})  we  conclude that   $\omega\pi=0$ and $P\omega =0$, that is we found two primary constraints.
Using the relations  (\ref{rel-1}) we find one more primary constraint,   $P\pi =0$. At last, computing $P^2+\pi^2$
given by (\ref{p1}) and (\ref{p2}) we see that all the terms with derivatives vanish, and we obtain the primary
constraint
\begin{equation}\label{p3}
T_1\equiv P^2 + m^2 c^2 +  \pi^2 - \frac{\alpha}{\omega^2}=0 \, .
\end{equation}
In the result, the action (\ref{L-curved}) implies four primary constraints,  $T_1$ and

\begin{eqnarray}\label{primary}
T_2\equiv\omega\pi=0, \quad T_3\equiv P\omega =0 , \quad T_4\equiv P\pi =0\,.
\end{eqnarray}
The  Hamiltonian is constructed excluding velocities from the expression
\begin{equation}\label{Hamiltonian-0}
H= p_\mu \dot x +\pi \dot\omega - L + \lambda_i T_i\equiv P\dot x + \pi\nabla\omega - L + \lambda_i T_i\, ,
\end{equation}
where $\lambda_i$ are the  Lagrangian multipliers associated with the primary constraints. From
(\ref{p1}) and (\ref{p2}), we observe the equalities $ P\dot x= (\sqrt{2}L_0)^{-1} (m^2c^2-
\frac{\alpha}{\omega^2})^{\frac 12} [\dot x N \dot x - \dot x K]$ and  $ \pi\nabla\omega =
(\sqrt{2}L_0)^{-1}(m^2c^2-\frac{\alpha}{\omega^2})^{\frac 12} [\nabla\omega N\nabla\omega  - \nabla\omega R]$. Together
with (\ref{rel-1})  they imply
$P\dot x + \pi\nabla\omega =L$.
Using this in (\ref{Hamiltonian-0}), we conclude that the Hamiltonian is composed from the primary constraints
\begin{equation}\label{Hamiltonian1}
H=\frac{\lambda_1}{2}\left( P^2 + m^2 c^2 + \pi^2 - \frac{\alpha}{\omega^2} \right) + \lambda_2 (\omega\pi) +\lambda_3
(P\omega) +\lambda_4 (P\pi )\, .
\end{equation}
The full set of phase-space coordinates consists of the pairs $x^\mu, p_\mu$ and  $\omega^\mu, \pi_\mu$.  They fulfill
the fundamental Poisson brackets $\{x^\mu ,  p_\nu\}=\delta^\mu_{\nu}$ and $\{\omega^\mu , \pi_\nu\}=\delta^\mu_\nu$,
then
$\quad \{P_\mu , P_\nu \} = R^\sigma_{\ \lambda \mu\nu} \pi_\sigma \omega^\lambda$, $\{P_\mu , \omega^\nu
\}=\Gamma^\nu_{\mu\alpha}\omega^\alpha$, $\{ P_\mu , \pi_\nu \}=- \Gamma^\alpha_{\mu\nu} \pi_\alpha$. For the
quantities $x^\mu$, $P^\mu$ and $S^{\mu\nu}$ these brackets imply the typical relations used by people for spinning
particles in Hamiltonian formalism
\begin{equation}\label{br}
\{ x^\mu , P_\nu \} =\delta^\mu_\nu,  \quad \{ P_\mu , P_\nu\}=-\frac14 R_{\mu\nu\alpha\beta}S^{\alpha\beta}, \quad \{
P_\mu, S^{\alpha\beta} \}=\Gamma^{\alpha}_{\mu\sigma}S^{\sigma\beta}-\Gamma^\beta_{\mu\sigma}S^{\sigma\alpha} \, .
\end{equation}
\begin{eqnarray}\label{br1}
\{S^{\mu\nu},S^{\alpha\beta}\}= 2(g^{\mu\alpha} S^{\nu\beta}-g^{\mu\beta} S^{\nu\alpha}-g^{\nu\alpha} S^{\mu\beta}
+g^{\nu\beta} S^{\mu\alpha})\,.
\end{eqnarray}
To reveal the higher-stage constraints and the Lagrangian multipliers we study the equations $\dot T_i=\{ T_i , H\}=0$.
$T_2$ implies the secondary constraint
\begin{equation}\label{secondary}
\dot T_2=0 \quad \Rightarrow \quad T_5\equiv \pi^2 - \frac{\alpha}{\omega^2}\approx 0,
\end{equation}
then  $T_1$ can be replaced on $P^2 + m^2c^2 \approx 0$. Preservation in time of $T_4$ and $T_3$ gives the Lagrangian
multipliers $\lambda_3$ and $\lambda_4$
\begin{equation}\label{FF3}
\lambda_3 =2a\lambda_1(\pi\theta P), \quad \lambda_4= -2a\lambda_1(\omega\theta P) \, ,
\end{equation}

where we have denoted
\begin{equation}\label{theta}
\theta_{\mu\nu} \equiv R_{\alpha\beta\mu\nu}S^{\alpha\beta} \, .
\end{equation}
\begin{eqnarray}\label{not.a33}
a=\frac{2}{16m^2c^2+(\theta S)} \,,
\end{eqnarray}
Preservation in time of $T_1$ gives the equation $\lambda_3(\omega\theta P) + \lambda_4(\pi\theta P) =0 $ which is
identically satisfied by virtue of (\ref{FF3}).  No more constraints are generated after this step . We summarize the
algebra of Poisson brackets between the constraints in the Table \ref{algebra-constraints}. $T_3$ and $T_4$ represent a
pair of second-class constraints, while $T_2$, $T_5$ and the  combination
\begin{equation} \label{T0}
T_0 = T_1 +4a(\pi\theta P) T_3 - 4a(\omega\theta P) T_4 \, ,
\end{equation}
are the first-class constraints. Taking into account that each second-class constraint rules out one phase-space
variable, whereas each first-class constraint rules out two variables, we have the right number of spin degrees of
freedom, $8-(2+4)=2$.

It should be noted that $\omega^\mu$ and $\pi^\mu$ turns out to be space-like vectors. Indeed, in flat limit and in the
frame where $p^\mu=(p^ 0,{\bf 0})$ the constraints $\omega p=\pi p=0$ imply $\omega^0=\pi^0=0$. This implies
$\omega^2\ge 0$ and $\pi^2\ge 0$. Combining this with the constraint (\ref{secondary}), we conclude $\omega^2>0$ and
$\pi^2>0$.

We point out that the first-class constraint $T_5=\pi^2 - \frac{\alpha}{\omega^2}\approx 0$ can be replaced on the pair
\begin{eqnarray}\label{primary0}
\pi^2=\mbox{const}, \qquad \omega^2=\mbox{const},
\end{eqnarray}
this gives an equivalent formulation of the model. The Lagrangian which implies the constraints (\ref{primary}) and
(\ref{primary0}) has been studied in \cite{DPW2, DPM2, DPM3}. Hamiltonian and Lagrangian equations for
physical variables of the two formulations coincide \cite{deriglazov2014Monster}, which proves their equivalence.

Using (\ref{FF3}), we can present the Hamiltonian (\ref{Hamiltonian1}) in the form
\begin{eqnarray}\label{Hamiltonian}
H= \frac{\lambda_1}{2} \left( P^2 + m^2c^2 +4a\left[(\pi\theta P)(P\omega)-(\omega\theta P)(P\pi )\right]
\right)+\frac{\lambda_1}{2}\left(  \pi^2 - \frac{\alpha}{\omega^2} \right)  + \lambda_2 (\omega\pi) \, .
\end{eqnarray}

\begin{table}
\caption{Algebra of constraints} \label{algebra-constraints}
\begin{center}
\begin{tabular}{|c|c|c|c|c|c|}
\hline
                              & $\qquad T_1 \qquad$  & $T_5$         & $T_2$          & $T_3$                 & $T_4$       \\  \hline \hline
$T_1=P^2+m^2 c^2$         & 0             & 0             & 0              & $\frac12 (\omega\theta P)$                    & $\frac12 (\pi\theta P)$     \\

                              &               &               &           &            &              \\ \hline
$ T_5=\pi^2 -\frac{\alpha}{\omega^2}    $               & 0             &   0   &  $-2T_5$             & $-2T_4$   & $-2T_3/\omega^2$   \\
& ${}$ &      &           &                &      \\
\hline
$T_2=\omega\pi$            & 0             & $2T_5$  & $0$            & $-T_3$          & $T_4$     \\
& ${}$ &      &           &                &        \\   \hline
$T_3=P\omega$               & $ -\frac12 (\omega\theta P)$    & $2T_4$      & $T_3$   &     0                 & $-\frac{1}{8a}$  \\
& ${}$ &      &           &                &        \\  \hline
$T_4=P \pi$       & $-\frac12 (\pi\theta P)$  &$2T_3/\omega^2$  & $-T_4$&   $\frac{1}{8a}$               & 0        \\
                              & ${}$ &      &           &                &     \\   \hline
\end{tabular}
\end{center}
\end{table}

The dynamics of basic variables is governed by Hamiltonian equations $\dot z=\{z , H\}$, where $z=(x, p, \omega, \pi)$,
and the Hamiltonian is given in (\ref{Hamiltonian}). Equivalently, we can use the first-order variational problem
equivalent to (\ref{L-curved})
\begin{eqnarray}\label{gm7}
S_H=\int d\tau ~ p_\mu\dot x^\mu+\pi_\mu\dot \omega^\mu- \left[\frac{\lambda_1}{2}\left( P^2 + (mc)^2 + \pi^2 -
\frac{\alpha}{\omega^2}\right)+\lambda_2 (\omega\pi) +\lambda_3 (P\omega)+\lambda_4  (P\pi)\right].
\end{eqnarray}
Variation with respect to $\lambda_i$ gives the constraints (\ref{p3}) and (\ref{primary}), while variation with
respect to $x, p, \omega$ and $\pi$ gives the dynamical equations.  By construction of $S_H$, the variational equation
$\frac{\delta S_{\kappa}}{\delta p_\mu}=0$ is equivalent to $\dot x^\mu=\{x^\mu, H\}$, and so on. The equations can be
written in a manifestly covariant form as follows:

\begin{eqnarray}
\dot x^\mu= \lambda_1 \left[ P^\mu + 2a[(\pi\theta P)\omega^\mu-(\omega\theta P)\pi^\mu] \right] \, ,
\qquad \qquad \qquad  \label{motion-x-1} \\
\nabla P_\mu =R^\alpha_{\ \ \beta\mu\nu} \pi_\alpha \omega^\beta  \dot x^\nu \, , \qquad \qquad \qquad
\qquad \qquad \qquad  \label{motion-P}  \\
\nabla\omega^\mu =-2\lambda_1 a(\omega\theta P) P^\mu + \lambda_2 \omega^\mu +\lambda_1\pi^\mu \, , \quad
\nabla\pi_\mu= -2\lambda_1 a(\pi\theta P) P_\mu- \lambda_2 \pi_\mu -\lambda_1\frac{\omega_\mu}{\omega^2} \, .
\label{motion-pi-omega}
\end{eqnarray}

According to general theory \cite{13, gitman1990quantization, deriglazov2010classical}, neither constraints nor
equations of motion do not determine the functions $\lambda_1$ and $\lambda_2$. Their presence in the equations of
motion implies that evolution of our basic variables is ambiguous. This is in correspondence with two local symmetries
presented in the model.  The variables with ambiguous dynamics do not represent observable quantities, so we need to
search  for variables that can be candidates for observables. Consider  the  antisymmetric tensor (\ref{FF.2}).
%
%
As a consequence of $T_3=0$ and $T_4=0$, this obeys the Pirani supplementary condition \cite{pirani:1956, Tulc, Dixon1964}

\begin{equation}
S^{\mu\nu}P_\nu = 0 \, .\label{condition1}
\end{equation}
Besides, the constraints $T_2$ and $T_5$ fix the value of square
\begin{eqnarray}\label{z1}
S^{\mu\nu} S_{\mu\nu} = 8\alpha, \label{condition2}
\end{eqnarray}
so we identify $S^{\mu\nu}$ with the Frenkel spin-tensor \cite{Frenkel}. The equations (\ref{condition1}) and
(\ref{condition2}) imply that only two components of spin-tensor are independent, as it should be for spin one-half
particle. Equations of motion for $S^{\mu\nu}$ follow from (\ref{motion-pi-omega}). Besides, we express equations
(\ref{motion-x-1}) and (\ref{motion-P}) in terms of the spin-tensor. This gives the system

\begin{eqnarray}
\dot x^\mu &=& \lambda_1\left[ P^\mu +a S^{\mu\beta}\theta_{\beta\alpha}P^\alpha\right] \, , \label{motion-x} \\
\nabla P_\mu &=&- \frac{1}{4} R_{\mu\nu\alpha\beta} S^{\alpha\beta}\dot  x^\nu\equiv-\frac{1}{4}\theta_{\mu\nu}\dot
x^\nu
\, ,  \label{motion-P-2.2} \\
\nabla S^{\mu\nu} &=& 2(P^\mu \dot x^\nu - P^\nu \dot x^\mu )\,, \label{motion-J.3}
\end{eqnarray}

where $\theta$ has been defined in (\ref{theta}). Eq. (\ref{motion-J.3}), contrary to the equations
(\ref{motion-pi-omega}) for $\omega$ and $\pi$, does not depend on $\lambda_2$. This proves that the spin-tensor is
invariant under local spin-plane symmetry. The remaining ambiguity due to  $\lambda_1$ is related with
reparametrization invariance and disappears when we work with physical dynamical variables $x^i(t)$. Equations
(\ref{motion-x})-(\ref{motion-J.3}) together with (\ref{condition1}) and (\ref{condition2}), form a closed system which
determines evolution of a spinning particle.

To obtain the Hamiltonian equations we can equally use the Dirac bracket constructed with help of second-class
constraints
\begin{equation}\label{DB}
\{A , B \}_D = \{A , B\} -\frac{1}{8a} \left[ \{ A , T_3 \} \{T_4 , B\} - \{ A , T_4\}\{ T_3 , B\} \right] \,.
\end{equation}
Since the Dirac bracket of a second-class constraint with any quantity vanishes, we can now omit $T_3$ and $T_4$ from
(\ref{Hamiltonian}), this yields the Hamiltonian
\begin{eqnarray}\label{Hamiltonian.0}
H_1= \frac{\lambda_1}{2} \left( P^2 + m^2c^2\right) +\frac{\lambda_1}{2}\left(  \pi^2 - \frac{\alpha}{\omega^2} \right)
+ \lambda_2 (\omega\pi) \,.
\end{eqnarray}
Then the equations (\ref{motion-x-1})-(\ref{motion-pi-omega}) can be obtained according the rule $\dot z=\{z, H_1\}_D$.
The quantities $x^\mu$, $P^\mu$ and $S^{\mu\nu}$, being invariant under spin-plane symmetry, have vanishing brackets
with the corresponding first-class constraints $T_2$ and $T_5$. So, obtaining equations for these quantities, we can
omit the last two terms in $H_1$, arriving at the familiar relativistic Hamiltonian

\begin{eqnarray}\label{Hamiltonian.0.02}
H_2= \frac{\lambda_1}{2} \left( P^2 + m^2c^2\right) \,.
\end{eqnarray}
The equations (\ref{motion-x})-(\ref{motion-J.3}) can be obtained according the rule $\dot z=\{z, H_2\}_D$. From
(\ref{Hamiltonian.0.02}) we conclude that our model describe spinning particle without gravimagnetic moment. The
Hamiltonian with gravimagnetic moment $\kappa$ have been proposed by Khriplovich \cite{Khriplovich1989} adding non
minimal interaction $\frac{\lambda_1}{2}\frac{\kappa}{16}R_{\mu\nu\alpha\beta}S^{\mu\nu}S^{\alpha\beta}$ to the
expression for $H_2$. The corresponding Lagrangian formulation will be constructed in section \ref{sec5.1}.

Let us exclude momenta $P^\mu$ and  the auxiliary variable $\lambda_1$ from the Hamiltonian equations. This yields second-order
equation for the particle's position $x^\mu(\tau)$. To achieve this, we
observe that the equation (\ref{motion-x}) is linear on $P$

\begin{equation}\label{xTP}
\dot x^\mu = \lambda_1  T^\mu_{\ \nu} P^\nu, \quad \textrm{with} \quad   T^\mu_{\ \nu}= \delta^\mu_\nu +
aS^{\mu\alpha}\theta_{\alpha \nu} \, .
\end{equation}
Using the identity
\begin{eqnarray}\label{Id0.0}
(S\theta S)^{\mu\nu}=-\frac{1}{2} (S\theta)S^{\mu\nu}, \quad \mbox{where} \quad
S\theta=S^{\alpha\beta}\theta_{\alpha\beta},
\end{eqnarray}
we find inverse of the matrix $T^\mu{}_\nu$
\begin{equation}\label{T-matrix}
\tilde T^\mu_{\ \ \nu} = \delta^\mu_{ \ \nu} - \frac{1}{8m^2c^2} S^{\mu\sigma}\theta_{\sigma\nu} \,, \qquad T^\mu_{\ \alpha}\tilde T^\alpha_{\  \nu}=\delta^\mu_\nu
\,,
\end{equation}
%
so (\ref{xTP}) can be solved with respect to $P^\mu$, $P^\mu=\frac{1}{\lambda_1}\tilde T^\mu{}_\nu\dot x^\nu$. We
substitute $P^\mu$ into the constraint $P^2+m^2c^2=0$, this gives expression for $\lambda_1$
\begin{equation}\label{L0}
\lambda_1= \frac{\sqrt {- G_{\mu\nu} \dot x^\mu \dot x^\nu}}{mc} \equiv \frac{\sqrt{-\dot x G \dot x}}{mc} \,.
\end{equation}
We have introduced the effective metric
\begin{equation}\label{G-metric}
G_{\mu\nu} \equiv \tilde T^\alpha_{\  \mu} g_{\alpha\beta}  \tilde T^{\beta}_{\ \nu} \, .
\end{equation}
The matrix $G$ is composed from the original metric $\eta_{\mu\nu}$ plus (spin and field-dependent) contribution,
$G_{\mu\nu}=\eta_{\mu\nu}+h_{\mu\nu}(S)$. So we call $G$ the effective metric produced along the world-line by
interaction of spin with gravity. The effective metric will play the central role in our discussion of
ultra-relativistic limit.

From (\ref{xTP}) and (\ref{L0}) we obtain the final expression for $P_\mu$
\begin{eqnarray}\label{PTx}
P^\mu =\frac{mc}{\sqrt{-\dot x G \dot x}} \tilde T^\mu_{\ \ \nu}\dot x^\nu =\frac{mc}{\sqrt{-\dot x G \dot x}}\left[
\dot x^\mu - \frac{1}{8m^2c^2} S^{\mu\nu}\theta_{\nu\sigma}\dot x^\sigma \right] \, ,
\end{eqnarray}
and Lagrangian form of the Pirani condition
\begin{eqnarray}\label{condition1a}
S^\mu{}_\nu\dot x^\nu-\frac{1}{8(mc)^2}(SS\theta\dot x)^\mu=0 \, . 
\end{eqnarray}
Using  the equations (\ref{PTx}) and (\ref{condition1a}) in (\ref{motion-P-2.2}) and (\ref{motion-J.3}) we finally
obtain
\begin{eqnarray}
\nabla\left[ \frac{ \tilde T^\mu_{\ \ \nu} \dot x^\nu}{\sqrt{-\dot xG\dot x}} \right] =- \frac{1}{4mc} R^\mu{}_{
\nu\alpha\beta}S^{\alpha\beta}\dot x^\nu \, , \label{byM13} \\
\nabla S^{\mu\nu}= \frac{1}{4mc\sqrt{-\dot xG\dot x}}\dot x^{[\mu} S^{\nu]\sigma} \theta_{\sigma\alpha}\dot x^\alpha \,
. \label{motionJ-5}
\end{eqnarray}

These equations, together with the conditions  (\ref{condition1a}) and (\ref{condition2}), form closed system for the
set ($x^\mu, S^{\mu\nu}$). The consistency of the constraints (\ref{condition1a}) and (\ref{condition2}) with the
dynamical equations is guaranteed by Dirac procedure for singular systems.

\subsection{Lagrangian action of spinning particle with unfixed value of spin}\label{sec4.2}

The Lagrangians (\ref{Lfree}) and (\ref{L-curved})  yield the fixed value of spin (\ref{z1}), that is they correspond
to an elementary particle. Let us present the modification which leads to the theory with unfixed spin, and, similarly
to Hanson-Regge approach \cite{hanson1974}, with a mass-spin trajectory constraint. Consider the following Lagrangian
in curved background
\begin{eqnarray}\label{aaa1}
L =-\frac{mc}{\sqrt{2}}\sqrt{-\dot x N \dot x - l^2\frac{\nabla\omega N \nabla\omega}{\omega^2} +\sqrt{\left[\dot x
N\dot x + l^2\frac{\nabla\omega N \dot\omega}{\omega^2}\right]^2- 4l^2\frac{(\dot x N \nabla\omega )^2}{\omega^2}}},
\end{eqnarray}
where $l$ is a parameter with the dimension of length. Applying the Dirac procedure as in Section \ref{sec4.1}, we obtain
the Hamiltonian
\begin{equation}\label{aaa2}
H=\frac{\lambda_1}{2}\left( P^2 + m^2 c^2 + \frac{\pi^2\omega^2}{l^2} \right) + \lambda_2 (\omega\pi) +\lambda_3
(P\omega) +\lambda_4 (P\pi )\,,
\end{equation}
which turns out to be combination of the first-class constraints $P^2 + m^2 c^2 + \frac{\pi^2\omega^2}{l^2}=0$,
$\omega\pi=0$ and the second-class constraints $P\omega=0$, $P\pi=0$. The Dirac procedure stops on the first stage,
that is there are no of secondary constraints. As compared with (\ref{L-curved}), the first-class constraint
$\pi^2-\frac{\alpha}{\omega^2}=0$ does not appear in the present model. Due to this, square of spin is not fixed,
$S^2=8(\omega^2\pi^2-\omega\pi)\approx8\omega^2\pi^2$. Using this equality, the mass-shell constraint acquires the
string-like form
\begin{equation}\label{aaa3}
P^2 + m^2 c^2 + \frac{1}{8l^2}S^2=0.
\end{equation}
The model has four physical degrees of freedom in the spin-sector. As the independent gauge-invariant degrees of
freedom, we can take three components $S^{ij}$ of the spin-tensor together with any one product of conjugate
coordinates, for instance, $\omega^0\pi^0$.

Using the auxiliary variable $\lambda$, we can rewrite the Lagrangian in the equivalent form
\begin{eqnarray}\label{aaa4}
L =\frac{1}{2\lambda}\left[\dot x N \dot x + l^2\frac{\nabla\omega N \nabla\omega}{\omega^2} -\sqrt{\left[\dot x
N\dot x + l^2\frac{\nabla\omega N \dot\omega}{\omega^2}\right]^2- 4l^2\frac{(\dot x N \nabla\omega )^2}{\omega^2}} \right]-\frac{\lambda}{4}m^2c^2.
\end{eqnarray}
Contrary to Eq. (\ref{aaa1}), it admits the massless limit.

\subsection{Mathisson-Papapetrou-Tulczyjew-Dixon (MPTD) equations and dynamics of representative point of a rotating body}\label{sec4.3}


In this section we discuss MPTD-equations of a rotating body in the form studied by Dixon\footnote{Our $S$ is twice of
that of Dixon.} (for the relation of the Dixon equations with those of Papapetrou and Tulczyjew see p. 335 in
\cite{Dixon1964})
\begin{eqnarray}
\nabla P^\mu=-\frac 14 R^\mu{}_{\nu\alpha\beta}S^{\alpha\beta}\dot x^\nu \equiv-\frac 14 (\theta\dot x)^\mu \, ,\label{r1}\\
\nabla S^{\mu\nu}= 2(P^\mu \dot x^\nu - P^\nu \dot x^\mu)\, , \label{r2} \\
S^{\mu\nu}P_\nu  =0, \label{r3}
\end{eqnarray}
and compare them with equations of motion of our spinning particle. In particular, we show that the effective metric
$G_{\mu\nu}$ also emerges in this formalism. MPTD-equations appeared in multipole approach to description of a body
\cite{Mathisson:1937zz, Papapetrou:1951pa, Tulc, Dixon1964, Dixon1965, Trautman2002}, where the energy-momentum of the
body is modelled by a set of multipoles. In this approach $x^\mu(\tau)$ is called representative point of the body, we
take it in arbitrary parametrization $\tau$ (contrary to Dixon, we do not assume the proper-time parametrization, that
is we do not add the equation $g_{\mu\nu}\dot x^\mu\dot x^\nu=-c^2$ to the system above). $S^{\mu\nu}(\tau)$ is
associated with inner angular momentum, and $P^\mu(\tau)$ is called momentum. The first-order equations (\ref{r1}) and
(\ref{r2}) appear in the pole-dipole approximation, while the algebraic equation (\ref{r3}) has been added by
hand\footnote{For geometric interpretation of the spin supplementary condition in the multipole approach see
\cite{Dixon1964}.}. After that, the number of equations coincides with the number of variables.

To compare MPTD-equations with those of section \ref{sec4.1}, we first observe some useful consequences of the system
(\ref{r1})-(\ref{r3}).

Take derivative of the constraint, $\nabla(S^{\mu\nu}P_\nu)=0$, and use (\ref{r1}) and (\ref{r2}), this gives the
expression
\begin{eqnarray}\label{r4}
(P\dot x)P^\mu=P^2\dot x^\mu+\frac18(S\theta \dot x)^\mu,
\end{eqnarray}
which can be written in the form
\begin{eqnarray}\label{r5}
P^\mu=\frac{P^2}{(P\dot x)}\left(\delta^\mu{}_\nu+\frac{1}{8P^2}(S\theta)^\mu{}_\nu\right)\dot x^\nu
\equiv\frac{P^2}{(P\dot x)}\tilde{\mathcal{T}}^\mu{}_\nu\dot x^\nu.
\end{eqnarray}
Contract (\ref{r4}) with $\dot x_\mu$. Taking into account that $(P\dot x)<0$, this gives
$(P\dot x)=-\sqrt{-P^2}\sqrt{-\dot x\tilde{\mathcal{T}}\dot x}$.
Using this in Eq. (\ref{r5}) we obtain
\begin{eqnarray}\label{r7}
P^\mu=\frac{\sqrt{-P^2}}{\sqrt{-\dot x\tilde T\dot x}}(\tilde{\mathcal{T}}\dot x)^\mu, \qquad \tilde
{\mathcal{T}}^\mu{}_\nu=\delta^\mu{}_\nu+\frac{1}{8P^2}(S\theta)^\mu{}_\nu.
\end{eqnarray}
For the latter use we observe that in our model with composite $S^{\mu\nu}$ we used the identity (\ref{Id0.0}) to
invert $T^\mu_{\ \nu}$, then the Hamiltonian equation (\ref{motion-x}) has been written in the form (\ref{PTx}), the latter can be
compared with (\ref{r7}).

Contracting  (\ref{r2}) with $S_{\mu\nu}$ and using  (\ref{r3}) we obtain $\frac{d}{d\tau}(S^{\mu\nu}S_{\mu\nu})=0$,
that is, square of spin is a constant of motion. Contraction of  (\ref{r4}) with $P_\mu$  gives $(PS\theta\dot x)=0$.
Contraction of (\ref{r4}) with $(\dot x\theta)_\mu$ gives $(P\theta\dot x)=0$. Contraction of (\ref{r1}) with $P_\mu$,
gives $\frac{d}{d\tau}(P^2)=-\frac12(P\theta\dot x)=0$, that is $P^2$ is one more constant of motion, say $k$,
$\sqrt{-P^2}=k=\mbox{const}$ (in our model this is fixed as $k=mc$). Substituting  (\ref{r7}) into the equations
(\ref{r1})-(\ref{r3}) we now can exclude $P^\mu$ from these equations, modulo to the constant of motion
$k=\sqrt{-P^2}$.

Thus, square of momentum can not be excluded from the system (\ref{r1})-(\ref{r4}), that is MPTD-equations in this form
do not represent a Hamiltonian system for the pair $x^\mu, P^\mu$. To improve this point, we note that Eq. (\ref{r7})
acquires a conventional form (as the expression for conjugate momenta of $x^\mu$ in the Hamiltonian formalism), if we
add to the system (\ref{r1})-(\ref{r3}) one more equation, which fixes the remaining quantity $P^2$ (Dixon noticed this
for the body in electromagnetic field, see his eq. (4.5) in \cite{Dixon1965}). To see, how the equation could look, we
note that for non-rotating body (pole approximation) we expect equations of motion of spinless particle, $\nabla
p^\mu=0$, $p^\mu=\frac{mc}{\sqrt{-\dot xg\dot x}}\dot x^\mu$, $p^2+(mc)^2=0$. Independent equations of the system
(\ref{r1})-(\ref{r4}) in this limit read $\nabla P^\mu=0$, $P^\mu=\frac{\sqrt{-P^2}}{\sqrt{-\dot xg\dot x}}\dot x^\mu$.
Comparing the two systems, we see that the missing equation is the mass-shell condition $P^2+(mc)^2=0$. Returning to
the pole-dipole approximation, an admissible equation should be $P^2+(mc)^2+f(S, \ldots)=0$, where $f$ must be a
constant of motion. Since the only constant of motion in arbitrary background is $S^2$, we have finally
\begin{eqnarray}\label{r9.1}
P^2=-(mc)^2-f(S^2).
\end{eqnarray}
With this value of $P^2$, we can exclude $P^\mu$ from MPTD-equations, obtaining closed system with second-order
equation for $x^\mu$ (so we refer the resulting equations as Lagrangian form of MPTD-equations). We substitute
(\ref{r7}) into (\ref{r1})-(\ref{r3}), this gives
\begin{eqnarray}
\nabla\frac{(\tilde{\mathcal{T}}\dot x)^\mu}{\sqrt{-\dot x\tilde{\mathcal{T}}\dot
x}}=-\frac{1}{4\sqrt{-P^2}}(\theta\dot x)^\mu,
\qquad \qquad \label{r9} \\
\nabla S^{\mu\nu}=
-\frac{1}{4\sqrt{-P^2}{\sqrt{-\dot x\tilde{\mathcal{T}}\dot x}}}\dot x^{[\mu}(S\theta\dot x)^{\nu]}, \label{r10} \\
(SS\theta\dot x)^\mu=-8P^2(S\dot x)^\mu, \qquad \qquad \qquad \label{r11}
\end{eqnarray}
where (\ref{r9.1}) is implied. They determine evolution of $x^\mu$ and $S^{\mu\nu}$ for each given function $f(S^2)$.

It is convenient to introduce the effective metric ${\mathcal{G}}$ composed from the "tetrad field"
$\tilde{\mathcal{T^\mu_{\  \nu}}}$
\begin{equation}\label{r12}
{\mathcal{G}}_{\mu\nu} \equiv g_{\alpha\beta} \tilde{\mathcal{T}}^{\alpha}{}_{\mu} \tilde{\mathcal{T}}^{\beta}{}_{\nu}
\, .
\end{equation}
Eq. (\ref{r11}) implies the identity
\begin{eqnarray}\label{r13}
\dot x\tilde{\mathcal{T}}\dot x=\dot x{\mathcal{G}}\dot x,
\end{eqnarray}
so we can replace $\sqrt{-\dot x\tilde{\mathcal{T}}\dot x}$ in (\ref{r9})-(\ref{r11}) by $\sqrt{-\dot
x{\mathcal{G}}\dot x}$.

In resume, we have presented MPTD-equations in the form
\begin{eqnarray}
P^\mu=\frac{\sqrt{-P^2}}{\sqrt{-\dot x{\mathcal{G}}\dot x}}(\tilde{\mathcal{T}}\dot x)^\mu \, , \quad  \nabla
P^\mu=-\frac 14 (\theta\dot x)^\mu \, , \quad \nabla S^{\mu\nu}= 2P^{[\mu} \dot x^{\nu]} \, , \quad S^{\mu\nu}P_\nu  =0
\, , \label{r003} \\
P^2+(mc)^2+f(S^2)=0 \, , \qquad \qquad \qquad \qquad \qquad \qquad  \label{r004} \\
S^2 \quad \mbox{is a constant of motion} \, , \qquad \qquad \qquad \qquad \qquad \quad  \label{r005}
\end{eqnarray}
with $\tilde{\mathcal{T^\mu_{\ \nu}}}$ given in (\ref{r7}). Now we are ready to compare them with Hamiltonian equations of our
spinning particle, which we write here in the form
\begin{eqnarray}
P^\mu =\frac{mc}{\sqrt{-\dot x G \dot x}}(\tilde T\dot x)^\mu, \quad \nabla P^\mu =-\frac 14 (\theta\dot x)^\mu \, ,
\quad \nabla S^{\mu\nu} = 2P^{[\mu} \dot x^{\nu]}\, , \quad
S^{\mu\nu}P_\nu  =0 \, , \label{m003} \\
P^2+(mc)^2=0 \, , \qquad \qquad \qquad \qquad \qquad \qquad \qquad \label{m004} \\
S^2=8\alpha \, , \qquad \qquad \qquad \qquad \qquad  \qquad \qquad \qquad \label{m005}
\end{eqnarray}
with $\tilde T^\mu_{\ \nu}$ given in (\ref{T-matrix}). Comparing the systems, we see that our spinning particle has fixed values of
spin and canonical momentum, while for MPTD-particle the spin is a constant of motion and momentum is a function of
spin. We conclude that all the trajectories of a body with given $m$ and $S^2=\beta$ are described by our spinning
particle with spin $\alpha=\frac{\beta}{8}$ and with the mass equal to $\sqrt{m^2-\frac{f^2(\beta)}{c^2}}$. In this
sense our spinning particle is equivalent to MPTD-particle.

We point out that our final conclusion remains true even if we do not add (\ref{r9.1}) to MPTD-equations: to study the
class of trajectories of a body with $\sqrt{-P^2}=k$ and $S^2=\beta$ we take our spinning particle with $m=\frac{k}{c}$
and $\alpha=\frac{\beta}{8}$.

MPTD-equations in the Lagrangian form (\ref{r9})-(\ref{r11}) can be compared with
(\ref{condition1a})-(\ref{motionJ-5}).

\subsection{Ultra-relativistic limit: the problems with MPTD-equations}\label{sec4.4}
The equations for trajectory (\ref{byM13}) and for precession of spin (\ref{motionJ-5}) became singular at critical
velocity which obeys the equation
\begin{eqnarray}\label{pe1}
\dot xG\dot x=0.
\end{eqnarray}
As we discussed in Introduction, the singularity determines behavior of the particle in
ultra-relativistic limit. In Eq. (\ref{pe1}) appeared the effective metric (\ref{G-metric}) instead of the original
metric $g_{\mu\nu}$. It should be noted that the incorporation of the constraints (\ref{primary}) and (\ref{secondary}) into
a variational problem, as well as the search for an interaction consistent with them represent very strong restrictions
on possible form of the Lagrangian. So, the appearance of effective metric seems to be unavoidable in a systematically
constructed model of spinning particle. The same conclusion follow from our analysis of MPTD-equations in section
\ref{sec4.3}.

The effective metric is composed from the original one plus (spin and field-dependent) contribution, $G=g+h(S)$. So we
need to decide, which of them the particle probes as the space-time metric. Let us consider separately the two
possibilities.

Let us use $g$ to define the three-dimensional geometry  (\ref{La.3.0})-(\ref{La.5}). This leads to two problems. The
first problem is that the critical speed turns out to be slightly more then the speed of light.  To see this, we use
the Pirani condition to write (\ref{pe1}) in the form
\begin{equation}\label{critical1}
-\left(\frac{dt}{dx^0}\right)^2\dot xG\dot x  = \left(c^2-{\bf v}\gamma{\bf v}\right)+\frac{1}{(2m^2c^2)^2}
\left(v\theta SS\theta v\right) =0 \, ,
\end{equation}
with $v^\mu$ defined in (\ref{La.5.1}). Using the expression $S^{\mu\nu}=2\omega^{[\mu}\pi^{\nu]}$, we obtain
\begin{equation}\label{critical2}
-\left(\frac{dt}{dx^0}\right)^2\dot xG\dot x = \left(c^2-{\bf v}\gamma{\bf v}\right)+ \frac{1}{(m^2c^2)^2}\left( \pi^2
(v\theta\omega)^2 + \omega^2(v\theta\pi)^2 \right) =0 \, .
\end{equation}
As $\pi$ and $\omega$ are space-like vectors (see the discussion below Eq. (\ref{T0}) ), the last term is non-negative,
this implies $|{\bf v}_{cr}|\ge c$. Let us confirm that generally this term is nonvanishing function of velocity, then
$|{\bf v}_{cr}|> c$. Assume the contrary, that this term vanishes at some velocity, then
\begin{eqnarray}\label{pe2}
v \theta \omega =\theta_{0i} \omega^i + \theta_{i0} v^i \omega^0 =0 \, ,\\
v \theta \pi = \theta_{0i} \pi^i +\theta_{i0} v^i \pi^0 =0 \, .
\end{eqnarray}
We analyze these equations in the following special case.  Consider a space with covariantly-constant curvature
$\nabla_\mu R_{\mu\nu\alpha\beta} =0$. Then $\frac{d}{d\tau}(\theta_{\mu\nu}S^{\mu\nu})=2\theta_{\mu\nu}\nabla
S^{\mu\nu}$, and using (\ref{motionJ-5}) we conclude that $\theta_{\mu\nu}S^{\mu\nu}$ is an integral of motion. We
further assume that the only non vanishing part is the electric \cite{Hartle1985} part of the curvature,
$R_{0i0j}=K_{ij}$, with $\det K_{ij} \ne 0$. Then the integral of motion acquires the form
\begin{eqnarray}\label{pe3}
\theta_{\mu\nu} S^{\mu\nu} = 2 K_{ij}S^{0i}S^{0j} \, .
\end{eqnarray}
Let us take the initial conditions for spin such that $K_{ij}S^{0i}S^{0j}\ne 0$, then this holds at any future instant.
Contrary to this,  the system (\ref{pe2}) implies $K_{ij}S^{0i}S^{0j}=0$. Thus, the critical speed does not always
coincide with the speed of light and, in general case, we expect that ${\bf v}_{cr}$ is both field and spin-dependent
quantity.

The second problem is that acceleration of MPTD-particle grows up in the ultra-relativistic limit. In the spinless
limit the equation (\ref{byM13}) turn into the geodesic equation. Spin causes deviations from the geodesic equation due
to right hand side of this equation, as well as due to the presence of the tetrad field $\tilde T^\mu{}_\nu$ and of the effective
metric $G$ in the left hand side. Due to the dependence of the tetrad field on the spin-tensor $S$, the singularity
presented in (\ref{motionJ-5}) causes the appearance of the term proportional to $\frac{1}{\sqrt{\dot xG\dot x}}$ in
the expression for longitudinal acceleration. In the result, the acceleration grows up to infinity as the particle's
speed approximates to the critical speed. To see this, we separate derivative of $\tilde T^\mu{}_\nu$ in  Eq. (\ref{byM13})
\begin{eqnarray}\label{pe4}
\nabla\left[ \frac{\dot x^\mu}{\sqrt{-\dot xG\dot x}} \right]=- T^\mu_{\ \alpha} \left( \nabla \tilde T^\alpha_{\
\beta} \right) \frac{\dot x^\beta}{\sqrt{-\dot xG\dot x}} -\frac{1}{4mc} T^\mu_{\ \nu} (\theta \dot x)^\nu \, .
\label{WD1}
\end{eqnarray}
Using (\ref{motionJ-5}) we obtain
\begin{equation}\label{pe5}
\left[ \nabla \tilde T^\mu{}_\nu\right] \dot x^\nu
=-\frac{S^{\mu\alpha}}{8m^2c^2}\left[\frac{R_{\alpha\nu\beta\sigma}\dot x^\beta (S\theta \dot x)^{\sigma}}{2mc
\sqrt{-\dot x G \dot x}}   +S^{\beta\sigma} (\nabla R_{\alpha\nu\beta\sigma}) \right] \dot x^\nu \, .
\end{equation}
Using this expression together with the identity $(TS)^{\mu\nu}=8m^2c^2a S^{\mu\nu}$, the equation (\ref{pe4}) reads
\begin{equation}\label{pe6}
\frac{d}{d\tau}\left[ \frac{\dot x^\mu}{\sqrt{-\dot xG\dot x}} \right] = \frac{f^\mu}{\sqrt{-\dot xG\dot x}} \, ,
\end{equation}
where we denoted
\begin{eqnarray}\label{pe7}
f^\mu \equiv aS^{\mu\alpha}\left[\frac{R_{\alpha\nu\beta\sigma}\dot x^\beta (S\theta \dot x)^{\sigma}}{2mc \sqrt{-\dot
x G \dot x}} +S^{\beta\sigma} (\nabla R_{\alpha\nu\beta\sigma}) \right] \dot x^\nu -\left( \Gamma  \dot x \dot x
\right)^\mu - \frac{\sqrt{-\dot xG\dot x}}{4mc} \left( T \theta \dot x \right)^\mu \,  .
\end{eqnarray}
It will be sufficient to consider static metric $g_{\mu\nu}({\bf x})$ with $g_{0i}=0$. Then three-dimensional metric
and velocity are
\begin{eqnarray}\label{pe7.1}
\gamma_{ij} = g_{ij}  , \quad v^i = \frac{c}{\sqrt{-g_{00}}} \frac{dx^i}{dx^0}  \, ,
\end{eqnarray}
Taking $\tau=x^0$, the spacial part of equation (\ref{pe6}) with this metric reads
\begin{equation} \label{pe8}
\left(\frac{dt}{dx^0}\right)^{-1} \frac{d}{dx^0} \left[ \frac{v^i}{\sqrt{-vGv}} \right] = \frac{f^i(v)}{\sqrt{-vGv}} \, .
\end{equation}
with $v^\mu$ defined in (\ref{La.5.1}), for the case
\begin{equation} \label{pe8.1}
v^\mu=(\frac{c}{\sqrt{-g_{00}}}, ~ {\bf v}),
\end{equation}
and
\begin{equation} \label{pe8.2}
-vGv=-v\tilde Tv=c^2-{\bf v}g{\bf v}+\frac{(vS\theta v)}{8m^2c^2}.
\end{equation}
In the result, we have presented the equation for trajectory in the form convenient for analysis of acceleration, see
(\ref{La.12}). Using  the definition of three-dimensional covariant derivative (\ref{La.8.2}), we present the
derivative on the l.h.s. of (\ref{pe8}) as follows
\begin{equation}\label{pe9}
\frac{d}{dx^0} \left[ \frac{v^i}{\sqrt{-vGv}} \right] = \frac{1}{\sqrt{-vGv}}\left[ {\mathcal M}^i_{\ k} \nabla_0 v^k -
\tilde\Gamma(\gamma) ^i_{jk}v^jv^k \frac{dt}{dx^0} +\frac{Kv^i}{2(-vGv)} \right] \, ,
\end{equation}
We have denoted
\begin{eqnarray}\label{pe9.1}
K=(\nabla_0 G_{\mu\nu}) v^\mu v^\nu - v^\mu G_{\mu 0}v^k \partial_k \ln{ (-g_{00}}), \qquad
{\mathcal M}^i{}_{k} = \delta^i{}_k -\frac{v^iv^\mu G_{\mu k}}{vGv}.
\end{eqnarray}
The matrix ${\mathcal M}^i{}_{k}$ has the inverse
\begin{equation}\label{pe10}
\tilde{\mathcal M}^i{}_{k} = \delta^i{}_k + \frac{ v^iv^\mu G_{\mu k}}{v^\sigma G_{\sigma 0} v^0} \,, \quad \mbox{then}
\quad \tilde{\mathcal M}^i{}_{k}v^k=v^i\frac{vGv}{v^\sigma G_{\sigma 0} v^0}.
\end{equation}
Combining these equations, we obtain the three-acceleration of our spinning particle
\begin{eqnarray}\label{pe11}
a^i=\left(\frac{dt}{dx^0}\right)^{-1} \nabla_0 v^i = \tilde{\mathcal M}^i_{\ k}\left[ f^k + (\tilde \Gamma v v)^k \right]+
\frac{ K v^i}{2v^\sigma G_{\sigma 0} } \, .
\end{eqnarray}
Finally, using manifest form of $f^i$ from (\ref{pe7}) we have
\begin{eqnarray}\label{pe11.1}
a^i= \frac{ \tilde{\mathcal M}^i{}_{k} \hat S^k }{\sqrt{-vGv}} -  c^2\tilde{\mathcal M}^i_{\
k}\frac{\gamma^{kj}\partial_j  g_{00}}{2g_{00}} -\frac{\sqrt{-vGv}}{4mc} \tilde{\mathcal M}^i_{\ k} (T\theta v)^k  +
\frac{Kv^i}{2v^\sigma G_{\sigma 0}} + a\tilde{\mathcal M}^i_{\ k} S^{k\alpha}  R_{\alpha\nu\beta\sigma;\lambda}
S^{\beta\sigma} v^\nu v^\lambda \, .
\end{eqnarray}
The longitudinal acceleration is obtained by projecting $a^i$ on the direction of velocity, that is
\begin{eqnarray}\label{pe12}
({\bf v}\gamma {\bf a}) =  \frac{ac ({\bf v}\gamma \tilde{\mathcal M})_k  \hat S^k}{2m\sqrt{-vGv}} - c^2 ({\bf v}\gamma
\tilde{\mathcal M})_k\frac{\gamma^{kj}\partial_j g_{00}}{2g_{00}} -\frac{\sqrt{-vGv}}{4mc}({\bf v}\gamma
\tilde{\mathcal M})_k (T\theta v)^k  +\cr \frac{K }{2v^\sigma G_{\sigma 0}} ({\bf v}\gamma{\bf v})+ a({\bf v}\gamma
\tilde{\mathcal M})_k S^{k\alpha}  R_{\alpha\nu\beta\sigma;\lambda} S^{\beta\sigma} v^\nu v^\lambda \, .
\end{eqnarray}
where  $\hat S^k = S^{k\mu}R_{\mu\nu\alpha\beta}v^\nu v^\alpha (S\theta v)^\beta$. As the speed of the particle closes
to the critical velocity, the longitudinal acceleration diverges due to the first term in (\ref{pe12}). In resume,
assuming that MPTD-particle sees the original geometry $g_{\mu\nu}$, we have a theory with unsatisfactory behavior in
the ultra-relativistic limit.

Let us consider the second possibility, that is we take $G_{\mu\nu}$ to construct the three-dimensional geometry
(\ref{La.3.0})-(\ref{La.5}).  With these definitions we have, by construction, $-\dot x G \dot x=
(\frac{dt}{dx^0})^2(c^2 - ({\bf v}\gamma{\bf v}))$, so the critical speed coincides with the speed of light. In the
present case, the expression for three-acceleration can be obtained in closed form for an arbitrary curved background.
Taking $\tau = x^0$ the spacial part of (\ref{pe6}) implies
\begin{eqnarray}\label{La.12-a}
\left(\frac{dt}{dx^0}\right)^{-1}\frac{d}{dx^0} \left[ \frac{v^i}{\sqrt{c^2-{\bf v}\gamma{\bf v}}}\right]
=\frac{f^i(v)}{\sqrt{c^2-{\bf v}\gamma{\bf v}}} \, .
\end{eqnarray}
where, from (\ref{pe7}), $f^i$ is given by
\begin{eqnarray}\label{pe7.0}
f^i \equiv aS^{i\alpha}\left[ \frac{R_{\alpha\nu\beta\sigma}v^\beta(S\theta v)^{\sigma}}{2mc\sqrt{c^2- {\bf
v}\gamma{\bf v}}} +S^{\beta\sigma} (\nabla R_{\alpha\nu\beta\sigma}) \right] v^\nu   -\Gamma^i{}_{\mu\nu}(G) v^\mu
v^l\nu - \frac{\sqrt{c^2- {\bf v}\gamma{\bf v}}}{4mc} \left( T \theta v \right)^i \,  .
\end{eqnarray}
Equation (\ref{La.12-a}) is of the form (\ref{La.12}), so the acceleration is given by (\ref{La.11}) and (\ref{La.20})
where, for the present case, $\gamma_{i j } = G_{ij} -\frac{G_{0i} G_{0j}}{G_{00}}$
\begin{eqnarray}\label{La.11-a}
a^i=\tilde M^i{}_j[f^j+\tilde\Gamma^j{}_{kl}(\gamma)v^kv^l]+\frac12\left(\frac{dt}{dx^0}\right)^{-1}\left[({\bf
v}\partial_0\gamma\gamma^{-1})^i-\frac{({\bf v}\partial_0\gamma{\bf v})}{c^2}v^i\right] \, ,
\end{eqnarray}
\begin{eqnarray}\label{La.20-a}
{\bf v}\gamma{\bf a}=\left(1-\frac{{\bf v}\gamma{\bf v}}{c^2}\right)\left[ ({\bf
v}\gamma)_i[f^i(v)+\tilde\Gamma^i{}_{kl}(\gamma)v^kv^l]+\frac12\left(\frac{dt}{dx^0}\right)^{-1}({\bf
v}\partial_0\gamma{\bf v})\right]\,.
\end{eqnarray}
With $f^i$ given in (\ref{pe7.0}), the longitudinal acceleration vanishes as $v\rightarrow c$.

Let us resume the results of this subsection. Assuming that spinning particle probes the three-dimensional space-time
geometry determined by the original metric $g$, we have a theory with unsatisfactory ultra-relativistic limit. First,
the critical speed, which the particle can not overcome during its evolution in gravitational field, can be more then
the speed of light. Second, the longitudinal acceleration grows up to infinity in the ultra-relativistic limit.
Assuming that the the particle sees the effective metric G(S) as the space-time metric, we avoided the two problems.
But the resulting theory still possess the problem.  The acceleration (\ref{La.11-a}) contains the singularity due to
$f^i\sim\frac{1}{\sqrt{c^2-({\bf v}\gamma{\bf v})}}$, that is at $v=c$ the acceleration becomes orthogonal to the
velocity, but remains divergent. We conclude that MPTD-equations do not seem promising candidate for description of a
relativistic rotating body.

\section{Non minimal interaction with gravitational field.}\label{sec5}

Can we modify the MPTD-equations to obtain a theory with reasonable behavior with respect to the original metric
$g_{\mu\nu}$? In the previous section we have noticed that the bad behavior of acceleration originates from the fact
that variation rate of spin (\ref{motionJ-5}) diverges in the ultra-relativistic limit, $\nabla
S\sim\frac{1}{\sqrt{\dot xG\dot x}}$, and contributes into the expression for acceleration (\ref{pe12}) through the
tetrad field $\tilde T^\mu_{\ \nu}(S)$. To improve this, we  remind that MPTD-equations result from minimal interaction of spinning
particle with gravitational field. In this section we  demonstrate that vector model of spin admits also a non-minimal
interaction which involves the interaction constant $\kappa$. By analogy with the
magnetic moment, the interaction constant $\kappa$ is called gravimagnetic moment \cite{Khriplovich1989}.
In the resulting theory, the equation for precession of spin, $\nabla S\sim\frac{1}{\sqrt{-\dot
xG\dot x}}$, is replaced by $\nabla S\sim\sqrt{-\dot xg\dot x}$. This improves the bad behavior of MPTD equations. As
it will be seen below, introducing $\kappa$ we effectively change the supplementary spin condition and hence the
definition of center of mass.

\subsection{Hamiltonian variational problem}\label{sec5.1}

We add the term
$\frac{\lambda_1}{2}\kappa R_{\alpha\beta\mu\nu}\omega^\alpha \pi^\beta \omega^\mu
\pi^\nu\equiv\frac{\lambda_1}{2}\frac{\kappa}{16}\theta S$ into the Hamiltonian action (\ref{gm7}). Thus we
consider the variational problem
\begin{eqnarray}\label{gmm1}
S_{\kappa}=\int d\tau ~ p_\mu\dot x^\mu+\pi_\mu\dot \omega^\mu-\left[\frac{\lambda_1}{2}\left( P^2+\kappa
R_{\alpha\beta\mu\nu}\omega^\alpha \pi^\beta \omega^\mu \pi^\nu + (mc)^2 + \pi^2 -
\frac{\alpha}{\omega^2}\right)+\lambda_2 (\omega\pi) +\lambda_3 (P\omega)+\lambda_4  (P\pi)\right].
\end{eqnarray}
on the space of independent variables $x^\mu, p_\nu$, $\omega^\mu, \pi_\nu$ and $\lambda_a$. We have denoted
$P_\mu\equiv p_\mu-\Gamma^\beta_{\alpha\mu}\omega^\alpha\pi_\beta$, $P^2=g^{\mu\nu}P_\mu P_\nu$ and so on.  Note also
that the first two terms can be identically rewritten in the general-covariant form $p_\mu\dot x^\mu+\pi_\mu\dot
\omega^\mu=P_\mu\dot x^\mu+\pi_\mu\nabla\omega^\mu$. Variation of the action with respect to $\lambda_a$ gives the
algebraic equations
\begin{eqnarray}\label{gmm2}
P^2+\kappa R_{\alpha\beta\mu\nu}\omega^\alpha \pi^\beta \omega^\mu \pi^\nu + (mc)^2 + \pi^2 -
\frac{\alpha}{\omega^2}=0,
\end{eqnarray}
\begin{eqnarray}\label{gmm3}
\omega\pi=0 , \qquad P\omega=0, \qquad  P\pi=0,
\end{eqnarray}
while variations with respect to the remaining variables yield dynamical equations which can be written in the
covariant form as follows\footnote{Note that by construction of $S_{\kappa}$, the variational equation $\frac{\delta
S_{\kappa}}{\delta p_\mu}=0$ is equivalent to $\dot x^\mu=\{x^\mu, H\}$, and so on.}
\begin{eqnarray}\label{gmm4}
\frac{\delta S_{\kappa}}{\delta p_\mu}=0 \quad \Leftrightarrow \quad \dot
x^\mu=\lambda_1P^\mu+\lambda_3\omega^\mu+\lambda_4\pi^\mu,
\end{eqnarray}
\begin{eqnarray}\label{gmm5}
\frac{\delta S_{\kappa}}{\delta x^\mu}=0 \quad \Leftrightarrow \quad \nabla P_\mu= -R_{\mu\nu\alpha\beta}\dot
x^\nu\omega^\alpha\pi^\beta-\frac12\lambda_1\kappa\nabla_\mu
R_{\sigma\nu\alpha\beta}\omega^\sigma\pi^\nu\omega^\alpha\pi^\beta,
\end{eqnarray}
\begin{eqnarray}\label{gmm6}
\frac{\delta S_{\kappa}}{\delta\pi_\mu}=0 \quad \Leftrightarrow \quad \nabla\omega^\mu=
\lambda_1\pi^\mu-\lambda_1\kappa R^\mu{}_{\alpha\beta\nu}\omega^\alpha\omega^\beta\pi^\nu+\lambda_2\omega^\mu+\lambda_4
P^\mu,
\end{eqnarray}
\begin{eqnarray}\label{gmm7}
\frac{\delta S_{\kappa}}{\delta\omega^\mu}=0 \quad \Leftrightarrow \quad \nabla\pi_\mu=
-\frac{\lambda_1\alpha}{\omega^4}\omega_\mu-\lambda_1\kappa
R_{\mu\nu\alpha\beta}\pi^\nu\omega^\alpha\pi^\beta-\lambda_2\pi_\mu-\lambda_3 P_\mu.
\end{eqnarray}
Eq. (\ref{gmm4}) has been repeatedly used to obtain the final form (\ref{gmm5})-(\ref{gmm7}) of the equations
$\frac{\delta S_{\kappa}}{\delta x^\mu}=0$, $\frac{\delta S_{\kappa}}{\delta\pi_\mu}=0$ and $\frac{\delta
S_{\kappa}}{\delta\omega^\mu}=0$. Computing time-derivative of the algebraic equations (\ref{gmm3}) and using
(\ref{gmm4})-(\ref{gmm7}) we obtain the consequences
\begin{eqnarray}\label{gmm8}
\pi^2 - \frac{\alpha}{\omega^2}=0,
\end{eqnarray}
\begin{eqnarray}\label{gmm9}
\lambda_3 = 4a\lambda_1\left[2(1-\kappa)R_{\alpha\beta\mu\nu}\omega^\alpha \pi^\beta \pi^\mu P^\nu
+\kappa\pi^\sigma(\nabla_\sigma R_{\mu\nu\alpha\beta}) \omega^\mu\pi^\nu\omega^\alpha\pi^\beta \right]   \, , \cr
\lambda_4 =-4a\lambda_1\left[2(1-\kappa)R_{\alpha\beta\mu\nu} \omega^\alpha \pi^\beta \omega^\mu P^\nu
+\kappa\omega^\sigma(\nabla_\sigma R_{\mu\nu\alpha\beta}) \omega^\mu\pi^\nu\omega^\alpha\pi^\beta \right] \, ,
\end{eqnarray}
Here and below we use the following notation. In the equation which relates velocity and momentum will appear the
matrix ${\mathcal T^\mu_{\ \nu}}$

\begin{equation}\label{gmm11}
{\mathcal T}^{\alpha}{}_{\nu} \equiv \delta^\alpha{}_\nu -(\kappa-1)aS^{\alpha \sigma}\theta_{\sigma\nu} \, , \qquad
a=\frac{2}{16m^2c^2+(\kappa+1)(S\theta)}\,.
\end{equation}
The matrix has an inverse given by
\begin{eqnarray}\label{gmm14}
\tilde {\mathcal T}^{\alpha}{}_{\nu} \equiv \delta^\alpha{}_\nu+(\kappa-1)bS^{\alpha \sigma}\theta_{\sigma\nu} \, ,
\qquad b=\frac{1}{8m^2c^2+\kappa(S\theta)} \,.
\end{eqnarray}
The vector $Z^\mu$ is defined by
\begin{eqnarray}\label{gmm12}
Z^\mu=\frac{b}{8c}S^{\mu\sigma}(\nabla_\sigma R_{\alpha\beta\rho\delta})S^{\alpha\beta}S^{\rho\delta}\equiv
\frac{b}{8c}S^{\mu\sigma}\nabla_\sigma(S\theta).
\end{eqnarray}
This vanishes in a space with homogeneous curvature, $\nabla R=0$.

The time-derivatives of (\ref{gmm2}), (\ref{gmm8}) and (\ref{gmm9}) do not yield new algebraic equations. Due to
(\ref{gmm8}) we can replace the constraint  (\ref{gmm2}) on $P^2+\kappa R_{\alpha\beta\mu\nu}\omega^\alpha\pi^\beta
\omega^\mu \pi^\nu + (mc)^2=0$. The obtained expressions for $\lambda_3$ and $\lambda_4$ can be used to exclude these
variables from the equations (\ref{gmm4})-(\ref{gmm7}).

The constraints and equations of motion do not determine the functions $\lambda_1$ and $\lambda_2$, that is the
non-minimal interaction preserves both reparametrization and spin-plane symmetries of the theory.  The presence of
$\lambda_1$ and $\lambda_2$ in the equations (\ref{gmm6}) and (\ref{gmm7}) implies that evolution of the basic
variables is ambiguous, so they are not observable.  To find the candidates for observables, we note once again that
(\ref{gmm6}) and (\ref{gmm7}) imply an equation for $S^{\mu\nu}$ which does not contain $\lambda_2$. So we rewrite
(\ref{gmm4}) and (\ref{gmm5}) in terms of spin-tensor and add to them the equation for $S^{\mu\nu}$, this gives the
system
\begin{eqnarray}
\dot x^\mu &=&\lambda_1\left[  {\mathcal T}^{\mu}{}_{\nu} P^\nu +\kappa \frac{ac}{b}Z^\mu\right] \, , \label{gmx-2}\\
\nabla P_\mu &=&- \frac{1}{4}\theta_{\mu\nu}\dot x^\nu-\frac{\lambda_1\kappa}{32}\nabla_\mu(S\theta)\, , \label{gmP-2} \\
\nabla S^{\mu\nu} &=&-\frac{\kappa\lambda_1}{4}(\theta S)^{[\mu\nu]}+2P^{[\mu}\dot x^{\nu]}   \, . \label{gmJ.3-2}
\end{eqnarray}
Besides, the constraints (\ref{gmm2}), (\ref{gmm3}) and (\ref{gmm8}) imply
\begin{eqnarray}
P^2+\frac{\kappa}{16}\theta S + (mc)^2=0, \label{gmm13} \\
S^{\mu\nu}P_\nu=0, \qquad S^{\mu\nu}S_{\mu\nu}=8\alpha \,.  \label{gmm13.1}
\end{eqnarray}
The equations (\ref{gmm13.1}) imply that only two components of spin-tensor are independent, as it should be for spin
one-half particle. Eq. (\ref{gmJ.3-2}), contrary to the equations for $\omega$ and $\pi$, does not depend on
$\lambda_2$. This proves that the spin-tensor is invariant under local spin-plane symmetry. The remaining ambiguity due
to  $\lambda_1$ is related with reparametrization invariance and disappears when we work with physical dynamical
variables $x^i(t)$. Equations (\ref{gmx-2})-(\ref{gmJ.3-2}), together with (\ref{gmm13}) and (\ref{gmm13.1}), form  a
closed system which determines evolution of the spinning particle.

The gravimagnetic moment $\kappa$ can generally take any value. When $\kappa=0$ we recover the  MPTD equations.

Let us exclude $P^\mu$ and $\lambda_1$ from the equations (\ref{gmP-2}) and (\ref{gmJ.3-2}). Using (\ref{gmm14}) we
solve (\ref{gmx-2}) with respect to $P^\mu$. Using the resulting expression in the constraint (\ref{gmm13}) we obtain
$\lambda_1$
\begin{eqnarray} \label{gmm15}
\lambda_1=\frac{\sqrt{-\dot x {\mathcal G}\dot x}}{m_r c}, \quad \mbox{with} \quad  m_r^2  \equiv m^2 +
\frac{\kappa}{16 c^2} (S\theta) -\kappa^2 Z^2 \, ,
\end{eqnarray}
where the effective metric now is  given by
\begin{eqnarray}\label{gmm16}
{\mathcal G}_{\mu\nu} = \tilde {\mathcal T}^\alpha{}_{\mu} g_{\alpha\beta} \tilde {\mathcal T}^\beta{}_{\nu} \,  .
\end{eqnarray}
Then the expression for momentum in terms of velocity implied by (\ref{gmx-2})  is
\begin{eqnarray}\label{gmm17}
P^\mu = \frac{m_r c}{\sqrt{-\dot x {\mathcal G} \dot x}} \tilde {\mathcal T}^{\mu}{}_{\nu} \dot x^\nu-\kappa c Z^\mu
\,.
\end{eqnarray}
We substitute this $P^\mu$ into (\ref{gmP-2}),  (\ref{gmJ.3-2})
\begin{eqnarray}
\nabla\left[ \frac{m_r }{\sqrt{-\dot x {\mathcal G} \dot x}} \tilde {\mathcal T}^\mu{}_\nu \dot x^\nu \right] =
-\frac{1}{4c} \theta^\mu{}_\nu\dot x^\nu  -  \kappa\frac{\sqrt{-\dot x{\mathcal G}\dot x}}{32m_r c^2}
\nabla^\mu(S\theta)+\kappa\nabla Z^\mu \, , \label{gml1} \\
\nabla S^{\mu\nu} =- \frac{\kappa\sqrt{-\dot x {\mathcal G} \dot x}}{4m_r c }(\theta S)^{[\mu\nu]} -\frac{2m_r c
(\kappa-1)b}{\sqrt{-\dot x{\mathcal G}\dot x}}\dot x^{[\mu} (S\theta\dot x)^{\nu]}  +2\kappa c\dot x^{[\mu}Z^{\nu]} \,
. \label{gml2}
\end{eqnarray}
Together with (\ref{gmm13.1}),  this gives us the Lagrangian equations for the spinning particle with gravimagnetic
moment.
Comparing our equations to those of spinning particle on electromagnetic background (see Eqs. (\ref{FF.6})-(\ref{FF.8})
below), we see that the two systems have the same structure after the identification $\kappa\sim\mu$ and
$\theta_{\mu\nu}\equiv R_{\mu\nu\alpha\beta}S^{\alpha\beta}\sim F_{\mu\nu}$, where $\mu$ is the magnetic moment. That
is a curvature influences trajectory of spinning particle in the same way as the electromagnetic field with the
strength $\theta_{\mu\nu}$.

\subsection{Ultra-relativistic limit requires the value of gravimagnetic moment $\kappa=1$}\label{sec5.2}

In the previous subsection we have formulated Hamiltonian variational problem for spinning particle with gravimagnetic
moment $\kappa$ in an arbitrary gravitational background. The model is consistent for any value of $\kappa$. When
$\kappa=0$, our equations of motion (\ref{gmx-2})-(\ref{gmJ.3-2}) coincide with MPTD-equations (\ref{r003}). As we have
shown above, they have unsatisfactory behavior in ultra-relativistic limit. Consider now our spinning particle with
gravimagnetic moment $\kappa=1$. This implies $\tilde{\mathcal T}^\mu{}_\nu=\delta^\mu{}_\nu$, ${\mathcal
G}_{\mu\nu}=g_{\mu\nu}$, and crucially simplifies the equations of motion\footnote{Besides $S^{\mu\nu}P_\nu=0$, there
are other supplementary spin conditions \cite{Papapetrou:1951pa, Tulc, Dixon1964, Dixon1965, pirani:1956}. In this
respect we point out that the MPTD theory implies this condition with certain $P_\nu$ written in Eq. (\ref{r003}).
Introducing $\kappa$, we effectively changed $P_\nu$ and hence changed the supplementary spin condition. For instance,
when $\kappa=1$ and in the space with $\nabla R=0$, we have $P^\mu=\frac{\tilde mc}{\sqrt{-\dot xg\dot x}}\dot x^\mu$
instead of (\ref{r003}).}. The Hamiltonian equations (\ref{gmx-2})-(\ref{gmJ.3-2}) read
\begin{eqnarray}
\frac{m_r c}{\sqrt{-\dot x g \dot x}}\dot x^\mu =P^\mu +cZ^\mu \, , \label{gmm18} \\
\nabla P_\mu =- \frac{1}{4}\theta_{\mu\nu}\dot x^\nu-\frac{\sqrt{-\dot x g \dot x}}{32m_r c}\nabla_\mu(S\theta) \, ,  \label{gmm19} \\
\nabla S^{\mu\nu} = -\frac{\sqrt{-\dot x g \dot x}}{4m_r c}(\theta S)^{[\mu\nu]}+2P^{[\mu} \dot x^{\nu]} \, ,
\label{gmm20}
\end{eqnarray}
while the Lagrangian equations are composed now by the equation for trajectory
\begin{eqnarray}\label{acc-5}
\nabla \left[ \frac{m_r\dot x^\mu}{\sqrt{-\dot x g \dot x}} \right] = -\frac{1}{4c} \theta ^\mu{}_{\nu}\dot x^\nu -
\frac{\sqrt{-\dot x g \dot x}}{32m_r c^2}\nabla^\mu(S\theta) +\nabla Z^\mu \, ,
\end{eqnarray}
and by the equation for precession of spin-tensor
\begin{equation}\label{motionS}
\nabla S^{\mu\nu} =-\frac{\sqrt{-\dot x g \dot x}}{4m_r c}(\theta S)^{[\mu\nu]}+2c\dot x^{[\mu}Z^{\nu]} \, .
\end{equation}
These equations can be compared with (\ref{byM13}) and (\ref{motionJ-5}).  In the modified theory:

\par \noindent 1. Time interval and distance should be unambiguously defined within the original space-time
metric $g_{\mu\nu}$. So the critical speed is equal to the speed of light.

\par \noindent 2. Covariant precession of spin (\ref{motionS}) has a smooth behavior, in particular, for
homogeneous field, $\nabla R=0$, we have $\nabla
S\sim \sqrt{-\dot x g\dot x}$ contrary to $\nabla S\sim\frac{1}{\sqrt{-\dot x g\dot x}}$ in the equation
(\ref{motionJ-5}).

\par \noindent 3. Even in homogeneous field  we have modified dynamics for both $x$ and $S$.
The equation (\ref{acc-5}) in the space with homogeneous curvature has the structure similar to (\ref{gm7.2}), hence we
expect that longitudinal acceleration vanishes as $v\rightarrow c$. Let us  confirm this by direct computations.

To find the acceleration, we separate derivative of the radiation mass $m_r$ and write equation (\ref{acc-5}) in the form
\begin{eqnarray}\label{acc-6}
\frac{d}{d\tau}\left[ \frac{\dot x ^\mu}{\sqrt{-\dot x g \dot x}} \right] = \frac{f^\mu}{\sqrt{-\dot x g \dot x}} \, ,
\end{eqnarray}
where the force is
\begin{eqnarray}\label{force-6}
f^\mu \equiv -\Gamma^\mu_{\alpha\beta}\dot x^\alpha \dot x^\beta  -\frac{\sqrt{-\dot x g \dot x}}{4m_r c} \theta^\mu_{\
\nu} \dot x^\nu+ \frac{\dot xg\dot x}{32m_r^2c^2}\nabla^\mu(S\theta)+\frac{\sqrt{-\dot xg\dot x}}{m_r}\nabla Z^\mu-\dot
x^\mu\frac{\dot m_r}{m_r}.
\end{eqnarray}
While this expression contains derivatives of spin due to $\dot m_r$-term, the resulting expression is non singular function of velocity
because  $\nabla S$ is a smooth function. Hence, contrary to Eq. (\ref{pe7}), the force now is non singular function of velocity.
We take $\tau=x^0$ in the spacial part of the system (\ref{acc-6}), this gives
\begin{equation}
\left(\frac{dt}{dx^0}\right)^{-1}\frac{d}{dx^0}\left[ \frac{v^i}{\sqrt{c^2-({\bf v}\gamma {\bf v})}}\right]
=\frac{f^i(v)}{\sqrt{c^2-({\bf v}\gamma {\bf v})}} \, ,
\end{equation}
where $f^i(v)$ is obtained from (\ref{force-6}) replacing $\dot x^\mu$ by $v^\mu$ of equation (\ref{La.5.1}).
This system is of the form (\ref{La.12}), so the acceleration is given by (\ref{La.11}) and (\ref{La.20})
\begin{eqnarray}\label{accc1}
a^i=\tilde M^i{}_j[f^j+\tilde\Gamma^j{}_{kl}(\gamma)v^kv^l]+\frac12\left(\frac{dt}{dx^0}\right)^{-1}\left[({\bf
v}\partial_0\gamma\gamma^{-1})^i-\frac{({\bf v}\partial_0\gamma{\bf v})}{c^2}v^i\right] \, ,
\end{eqnarray}
\begin{eqnarray}\label{accc2}
{\bf v}\gamma{\bf a}=\left(1-\frac{{\bf v}\gamma{\bf v}}{c^2}\right)\left[
({\bf v}\gamma)_i[f^i(v)+\tilde\Gamma^i{}_{kl}(\gamma)v^kv^l]+\frac12\left(\frac{dt}{dx^0}\right)^{-1}({\bf v}\partial_0\gamma{\bf v})\right]\,.
\end{eqnarray}
With the smooth $f^i$ given in Eq. (\ref{force-6}), and as $v\rightarrow c$, the acceleration (\ref{accc1}) remains
finite while the longitudinal acceleration (\ref{accc2}) vanishes. Due to the identity (\ref{La.17.2}), we have $({\bf
v}\gamma)_if^i\stackrel{v\rightarrow c}{\longrightarrow}-_{\bf v}\gamma)_i\Gamma^i{}_{\alpha\beta}\dot x^\alpha \dot
x^\beta$, that is  the trajectory tends to that of spinless particle in the limit.

In resume, contrary to MPTD-equations, the modified theory is consistent with respect to the original metric
$g_{\mu\nu}$. Hence the modified equations could be more promising for description of the rotating objects in
astrophysics. It would be interesting to see if the non-minimal interaction allows to remove the spacelike-timelike
transitions observed for spinning particle in the Schwarzschild background \cite{Koch2015}.

We should note that MPTD-equations follow from a particular form assumed for the multipole representation of a rotating
body \cite{Trautman2002}. It would be interesting to find a set of multipoles which yields the modified equations
(\ref{gml1}) and (\ref{gml2}).

\subsection{Lagrangian of spinning particle with gravimagnetic moment}\label{sec5.3}

We look for the Lagrangian which in the phase space implies the variational problem (\ref{gmm1}). First, we note that
the constraints $\omega\pi=P\omega=0$ always appear from the Lagrangian which depends on $N\dot x$ and $N\dot\omega$. So we omit the corresponding terms in (\ref{gmm1}).
Second, we present the remaining terms in (\ref{gmm1}) in the form
\begin{eqnarray}\label{acc-7}
S_{\kappa}=\int d\tau ~ p_\mu\dot x^\mu+\pi_\mu\dot \omega^\mu-\frac{\lambda_1}{2}(P, \pi)\left(
\begin{array}{cc}
g& \lambda g\\
\lambda g& \sigma
\end{array}
\right)\left(
\begin{array}{c}
P\\
\pi
\end{array}
\right)-\frac{\lambda_1}{2}\left[(mc)^2-\frac{\alpha}{\omega^2}\right],
\end{eqnarray}
where we have introduced the symmetric matrix
\begin{eqnarray}\label{acc-8}
\sigma^{\mu\nu}=g^{\mu\nu}+\kappa R_\alpha{}^\mu{}_\beta{}^\nu\omega^\alpha\omega^\beta, \quad \mbox{then} \quad
\sigma^{\mu\nu}\omega_\nu=\omega^\mu.
\end{eqnarray}
The matrix appeared in (\ref{acc-7}) is invertible, the inverse matrix is
\begin{eqnarray}\label{acc-9}
\left(
\begin{array}{cc}
K\sigma& -\lambda K\\
-\lambda K& K
\end{array}\right) \,, \quad \mbox{where} \quad K=(\sigma-\lambda^2g)^{-1}.
\end{eqnarray}
When $\kappa=0$ we have $K^{\mu\nu}=(1-\lambda^2)^{-1}g^{\mu\nu}$, and (\ref{acc-9}) coincides with the matrix appeared
in (\ref{FF.1.2}). Third, we remind that the Hamiltonian variational problem of the form $p\dot
q-\frac{\lambda_1}{2}pAp$ follows from the reparametrization-invariant Lagrangian $\sqrt{\dot qA^{-1}\dot q}$. So, we
tentatively replace the matrix appeared in the free Lagrangian (\ref{FF.1.2}) by (\ref{acc-9}) and switch on the
minimal interaction of spin with gravity, $\dot\omega\rightarrow\nabla\omega$. This gives the following Lagrangian
formulation of spinning particle with gravimagnetic moment:
\begin{eqnarray}
L=-\sqrt{(mc)^2 -\frac{\alpha}{\omega^2}} ~ \sqrt{-(N\dot x, N\nabla\omega)\left(
\begin{array}{cc}
K\sigma& -\lambda K\\
-\lambda K& K
\end{array}
\right)\left(
\begin{array}{c}
N\dot x\\
N\nabla\omega
\end{array}
\right)}=  \label{acc-10}  \\
-\sqrt{(mc)^2 -\frac{\alpha}{\omega^2}} ~ \sqrt{-\dot xNK\sigma N\dot x-\nabla\omega NKN\nabla\omega+2\lambda\dot
xNKN\nabla\omega} \,.  \label{acc-11}
\end{eqnarray}

Let us show that this leads to the desired Hamiltonian formulation (\ref{gmm1}). The matrixes $\sigma$, $K$ and $N$ are
symmetric and mutually commuting. Canonical momentum for $\lambda$ vanishes and hence represents the primary
constraint, $p_\lambda=0$. In terms of the canonical momentum $P_\mu\equiv
p_\mu-\Gamma^\beta_{\alpha\mu}\omega^\alpha\pi_\beta$, the expressions for conjugate momenta $p_\mu=\frac{\partial
L}{\partial\dot x^\mu}$ and  $\pi_\mu=\frac{\partial L}{\partial\dot\omega^\mu}$ read
\begin{eqnarray}\label{acc-12}
P_\mu& =& \frac{1}{L_0}\left[m^2c^2 -\frac{\alpha}{\omega^2}\right]^{\frac12}\left[(\dot xNK\sigma
N)_\mu-\lambda(\nabla\omega NKN)_\mu\right],
\end{eqnarray}
\begin{eqnarray}\label{acc-13}
\pi_\mu &=& \frac{1}{\sqrt{2}L_0}\left[m^2c^2 -\frac{\alpha}{\omega^2}\right]^{\frac12} \left[(\nabla\omega
NKN)_\mu-\lambda(\dot xNKN)_\mu\right],
\end{eqnarray}
where $L_0$ is the second square root in (\ref{acc-11}). They immediately imply the primary constraints $\omega\pi=0$
and $P\omega=0$. From the expressions
\begin{eqnarray}\label{acc-14}
P^2=\frac{1}{L_0^2}\left[(mc)^2 -\frac{\alpha}{\omega^2}\right]\left[(\dot xNK\sigma K\sigma N\dot
x)+\lambda^2(\nabla\omega NKKN\nabla\omega)-2\lambda(\dot xNK\sigma KN\nabla\omega)\right], \cr
\pi\sigma\pi=\frac{1}{L_0^2}\left[(mc)^2 -\frac{\alpha}{\omega^2}\right]\left[\lambda^2(\dot xNK\sigma K N\dot
x)+(\nabla\omega NK\sigma KN\nabla\omega)-2\lambda(\dot xNK\sigma KN\nabla\omega)\right], \cr 2\lambda
P\pi=\frac{1}{L_0^2}\left[(mc)^2 -\frac{\alpha}{\omega^2}\right]\left[-2\lambda^2(\dot xNK\sigma K N\dot
x)-2\lambda^2(\nabla\omega NK KN\nabla\omega)+2\lambda(\dot xNK\sigma KN\nabla\omega)+ \right. \cr  \left.
2\lambda^3(\dot xNKKN\nabla\omega)\right], \qquad \qquad \qquad \qquad \qquad \qquad
\end{eqnarray}
we conclude that their sum does not depend on velocities and hence gives one more constraint
\begin{eqnarray}\label{acc-15}
P^2+\pi\sigma\pi+2\lambda P\pi=-\left[(mc)^2 -\frac{\alpha}{\omega^2}\right].
\end{eqnarray}
Then the Hamiltonian is
$H=P\dot x +\pi \nabla\omega - L + \lambda_i T_i$.
From (\ref{acc-12}) and (\ref{acc-13}) we obtain $P\dot x +\pi \nabla\omega=L$, so the Hamiltonian is composed from
primary constraints
\begin{equation}\label{acc-16}
H=\frac{\lambda_1}{2}\left[P^2+\kappa
R_{\alpha\mu\beta\nu}\omega^\alpha\pi^\mu\omega^\beta\pi^\nu+(mc)^2+\pi^2-\frac{\alpha}{\omega^2}+
2\lambda(P\pi)\right]+\lambda_2(\omega\pi)+\lambda_3(P\omega)+vp_\lambda \,.
\end{equation}
This expression is equivalent to the Hamiltonian written in the variational problem (\ref{gmm1}). The problems
(\ref{gmm1}) and (\ref{acc-16}) yield the same equations for the set $x^\mu$, $P^\mu$ and $S^{\mu\nu}$.

\section{Spinless particle non-minimally interacting with electromagnetic field, speed of light and critical speed}\label{sec6}

In this section we consider examples of manifestly Poincare-covariant and reparametrization-invariant equations of
spinless particle in Minkowski space which lead to critical speed different from the speed of light.
We achieve this assuming a non-minimal interaction with electromagnetic field.
These toy models confirm that the critical speed different from the speed of light
does not contradict relativistic invariance (existence of the observer-independent scale $c$).

Let us denote $x^i(t)$, $i=1, 2, 3$ physical dynamical variables describing trajectory of relativistic particle subject
to an external force. In order to work with manifestly covariant quantities, we use parametric equations of the
trajectory
$x^\mu(\tau)=(x^0\equiv ct(\tau), ~  x^i(\tau))$,
where $\tau$ is an arbitrary parameter along the world line. In this section we use the usual special-relativity notions for time,
three-dimensional distance, velocity and acceleration as well as for the scalar product
\begin{eqnarray}\label{L.1.1}
v^i=\frac{dx^i}{dt}, \qquad a^i=\frac{dv^i}{dt}, \qquad {\bf v}{\bf a}=v^ia^i.
\end{eqnarray}
Let us start from the standard Lagrangian of spinless particle in electromagnetic field. Using the auxiliary variable $\lambda$, the Lagrangian reads
\begin{eqnarray}\label{LL.0}
L=\frac{1}{2\lambda}\dot x^2-\frac{\lambda}{2}m^2c^2+\frac{e}{c}A\dot x.
\end{eqnarray}
This implies the manifestly relativistic equations
\begin{eqnarray}\label{LL.1}
\left(\frac{\dot x^i}{\sqrt{-\dot x^2}}\right)^.=\frac{e}{mc^2}F^\mu{}_\nu \dot x^\nu.
\end{eqnarray}
They became singular as $\dot x^2\rightarrow 0$.
Using the reparametrization-invariant derivative $D=\frac{1}{\sqrt{-\dot x^2}}\frac{d}{d\tau}$, they read in manifestly
reparametrization-invariant form
\begin{eqnarray}\label{LL.2}
DDx^\mu=f^\mu\equiv\frac{e}{mc^2}F^\mu{}_\nu Dx^\nu.
\end{eqnarray}
Due to the identities
\begin{eqnarray}\label{LL.2.1}
\dot x_\mu DDx^\mu=0, \qquad \dot x_\mu f^\mu=0,
\end{eqnarray}
the system contains only three independent equations. The first identity became more
transparent if we compute
derivative on the left hand side of (\ref{LL.1}), then the system reads
\begin{eqnarray}\label{LL.2.2}
M^\mu{}_\nu\ddot x^ \nu=\frac{e\sqrt{-\dot x^2}}{mc^2}F^\mu{}_\nu\dot x^\nu,
\end{eqnarray}
where $M^\mu{}_\nu$ turns out to be projector on the plane orthogonal to $\dot x^\mu$
\begin{eqnarray}\label{LL.2.3}
M^\mu{}_\nu=\delta^\mu{}_\nu-\frac{\dot x^\mu\dot x_\nu}{\dot x^2}, \qquad \dot x_\mu M^\mu{}_\nu=0.
\end{eqnarray}
Using reparametrization invariance, we can take physical time as the parameter, $\tau=t$,
this directly yields
equations for observable dynamical variables $x^ i(t)$.  In the physical-time parametrization we have $x^\mu=(ct, {\bf
x}(t))$, $\dot x^\mu=(c, {\bf v}(t))$ and $\frac{1}{\sqrt{-\dot x^2}}=\frac{1}{\sqrt{c^2-{\bf v}^2}}$. Time-like
component of the system (\ref{LL.1}) reads
\begin{eqnarray}\label{LL.3}
\frac{d}{dt}\frac{c}{\sqrt{c^2-{\bf v}^2}}=\frac{e}{mc^2}F^0{}_i \dot x^i.
\end{eqnarray}
and gives the value of acceleration along the direction of velocity
\begin{eqnarray}\label{LL.4}
{\bf v} {\bf a}=\frac{e(c^2-{\bf v}^2)^{\frac{3}{2}}}{mc^3}({\bf E}{\bf v}).
\end{eqnarray}
The longitudinal acceleration vanishes as $|{\bf v}|\rightarrow c$. Hence the singularity in Eq. (\ref{LL.1}) implies
that the particles speed can not exceed the value $c$.

Components of three-acceleration vector can be obtained from the spatial part of the system (\ref{LL.1})
\begin{eqnarray}\label{LL.5}
\frac{\ddot x^i}{\sqrt{c^2-{\bf v}^2}}+\dot x^i\frac{d}{dt}\frac{1}{\sqrt{c^2-{\bf v}^2}}=\frac{e}{mc^2}F^i{}_\nu \dot x^\nu.
\end{eqnarray}
Using Eq. (\ref{LL.3}) we obtain
\begin{eqnarray}\label{LL.6}
{\bf a}=\frac{\sqrt{c^2-{\bf v}^2}}{mc}\left[e{\bf E}-\frac{e({\bf E}{\bf v})}{c^2}{\bf v}+\frac{e}{c}{\bf v}\times{\bf
B}\right].
\end{eqnarray}
In accordance with degeneracy (\ref{LL.2.1}), scalar product of the spacial part
with ${\bf v}$ gives the time-like
component, Eq. (\ref{LL.4}).

Let us discuss two modifications which preserve both relativistic covariance and reparametrization invariance of
equations of motion, but could yield non vanishing longitudinal acceleration as $|{\bf v}|\rightarrow c$.

First, in the presence of external fields we can construct an additional reparametrization invariants. For instance, we
can use the derivative
\begin{eqnarray}\label{L.6.1}
D'\equiv\frac{1}{\sqrt{-\dot xg\dot x}}\frac{d}{d\tau},
\end{eqnarray}
where the usual relativistic factor $\dot x\eta\dot x$ is replaced by
\begin{eqnarray}\label{L.6.2}
-\dot x^\mu g_{\mu\nu}\dot x^\nu=-\dot x^2-\epsilon k(\dot xFF\dot x)= c^2-{\bf v}^2-\epsilon k(\dot xFF\dot x), \qquad \epsilon=\pm 1.
\end{eqnarray}
The right dimension can be supplied by the constant $k$ equal to $\frac{e^6}{m^4c^8}$ or $\frac{\hbar^3}{m^4c^5}$.
Second, we consider non minimal interaction with the force constructed from reparametrization-invariant quantities,
$f^\mu(D'x, F, \partial F, \ldots)=f^\mu{}_0+f^\mu{}_\nu D'x^\nu+\ldots$. Hence let us consider the manifestly
covariant and reparametrization-invariant equations
\begin{eqnarray}\label{LL.6}
D'D'x^\mu=f^\mu, \quad {\mbox or} \quad M^\mu{}_\nu\ddot x^\nu=-(\dot xg\dot x)f^\mu+\frac{(\dot x\dot g\dot x)}{2(\dot xg\dot x)}\dot x^\mu,
\end{eqnarray}
where
\begin{eqnarray}\label{LL.6.1}
M^\mu{}_\nu=\delta^\mu{}_\nu-\frac{\dot x^\mu(\dot xg)_\nu}{(\dot xg\dot x)}, \quad (\dot xg)_\mu M^\mu{}_\nu=0,
\quad M^\mu{}_\nu\dot x^ \nu=0.
\end{eqnarray}
Due to non invertibility of $M^\mu{}_\nu$, the system (\ref{LL.6}) consist of three second-order equations and one
equation of first order. Contracting (\ref{LL.6}) with $(\dot xg)_\mu$, we obtain the first-order equation
\begin{eqnarray}\label{LL.6.2}
\dot x\dot g\dot x=2(\dot xg\dot x)(\dot xgf).
\end{eqnarray}
This is analog of $\dot x_\mu f^\mu=0$ of the previous case (\ref{LL.2.1}), and can be used to present $f^0$ through
three-dimensional force. Using reparametrization invariance, we take $\tau=x^0$ and write (\ref{LL.6}) in the form
\begin{eqnarray}\label{LL.6.3}
\frac{d}{dx^0}\frac{1}{\sqrt{-\dot xg\dot x}}=\sqrt{-\dot xg\dot x}f^0,
\end{eqnarray}
\begin{eqnarray}\label{LL.6.4}
\frac{\ddot x^i}{\sqrt{-\dot xg\dot x}}+\dot x^i\frac{d}{dx^0}\frac{1}{\sqrt{-\dot xg\dot x}}=\sqrt{-\dot xg\dot x}
f^i.
\end{eqnarray}
Using (\ref{LL.6.3}) in (\ref{LL.6.4}) we obtain
\begin{eqnarray}\label{LL.6.5.0}
\frac{d^2 x^i}{dx^{02}}=-\dot xg\dot x\left[f^i-f^0 \frac{dx^i}{dx^0}\right], \qquad \dot x^\mu=\frac{dx^\mu}{dx^0}.
\end{eqnarray}
We take $x^0=ct$, then three-acceleration
\begin{eqnarray}\label{LL.6.5}
{\bf a}=-\dot xg\dot x\left[{\bf f}-\frac{f^0}{c}{\bf v}\right], \qquad \dot x^\mu=(c, {\bf v}=\frac{d{\bf x}}{dt}).
\end{eqnarray}
Equations (\ref{LL.6.2}), (\ref{LL.6.5}) are equivalent to the initial system (\ref{LL.6}). Eq. (\ref{LL.6.5}) implies
the longitudinal acceleration
\begin{eqnarray}\label{LL.6.6}
{\bf v} {\bf a}=-\dot xg\dot x\left[{\bf v} {\bf f}-\frac{{\bf v}^2}{c}f^0\right]\equiv -\dot xg\dot
x\left[\frac{c^2-{\bf v}^2}{c}f^0+\dot x_\mu f^\mu\right], \qquad f^\mu=f^\mu(\frac{\dot x}{\sqrt{-\dot xg\dot x}}, F,
\partial F, \ldots).
\end{eqnarray}
The acceleration vanishes at the values of speed which zero out r.h.s. of this equation. If in physical-time
parametrization the four-force obeys the identity $\dot x_\mu f^\mu=0$, we have two special points, $|{\bf v}|=c$, and
$v'$ determined from $\dot xg\dot x=0$. In absence of the  identity, and if $\dot x_\mu f^\mu\ne 0$ as $|{\bf
v}|\rightarrow c$, the speed of light does not represent a special point of (\ref{LL.6.6}).  Let us illustrate this
equation  with two examples. \par\noindent {\bf Example 1.} Consider the minimal interaction
$f^\mu=\frac{e}{mc^2}F^\mu{}_\nu D' x^\nu$ and the relativistic-contraction factor (\ref{L.6.2}), then (\ref{LL.6.6})
reads
\begin{eqnarray}\label{LL.6.7}
{\bf v} {\bf a}=\frac{e({\bf v} {\bf E})}{mc^3}\sqrt{-\dot xg\dot x}(c^2-{\bf v}^2).
\end{eqnarray}
Besides the usual special point, ${\bf v}^2=c^2$, there is one more, say $v'=|{\bf v}'|$, determined by $\dot xg\dot
x=0$. This surface is slightly different from the sphere $c^2-{\bf v}^2=0$. So the second special point generally
differs from the speed of light. To see this, we compute the last term in (\ref{L.6.2})
\begin{eqnarray}\label{L.6.4}
-\dot xFF\dot x=c^2E_i\left(\delta_{ij}-\frac{v_iv_j}{c^2}\right)E_j+{\bf v}^2B_iN_{ij}B_j.
\end{eqnarray}
Here $N_{ij}\equiv\delta_{ij}-\frac{v_iv_j}{{\bf v}^2}$ is projection operator on the plane orthogonal to the vector $\bf v$,
so we can write ${\bf B}N{\bf B}=(N{\bf B})^2={\bf B}_{\perp}^2$. Then the factor (\ref{L.6.2}) reads
\begin{eqnarray}\label{L.6.5}
-\dot xg\dot x=c^2-{\bf v}^2+\epsilon k\left[c^2{\bf E}(1-\frac{{\bf v}{\bf v}}{c^2}){\bf E}+{\bf v}^2{\bf
B}_{\perp}^2\right].
\end{eqnarray}
The quantity $\delta_{ij}-\frac{v_iv_j}{c^2}$ turns into the projection operator $N$ when ${|\bf v|}=c$. Hence
\begin{eqnarray}\label{L.6.6}
-\dot xg\dot x\stackrel{{|\bf v|}\rightarrow c}{\longrightarrow}\epsilon kc^2[{\bf E}_{\perp}^2+{\bf B}_{\perp}^2].
\end{eqnarray}
If ${\bf E}$ and ${\bf B}$ are not mutually parallel in the laboratory system, this expression does not vanish for any
orientation of ${\bf v}$.  This implies that the factor (\ref{L.6.2}) does not vanish at ${|\bf v|}=c$.

We confirmed that  longitudinal acceleration generally vanishes at two different values of speed, $c$ and $v'$. Then
Eq. (\ref{LL.6.7}) implies the following possibilities.

\par\noindent A) Let $\epsilon=+1$, then from
(\ref{L.6.5}) we conclude $c<v'$, and speed of the particle approximates to $c$.  The second special point $v'$ turns
out to be irrelevant. So $v_{cr}=c$.

\par\noindent B) Let $\epsilon=-1$, then $v'<c$, and the particle with small initial
velocity will approximate to the critical velocity $v_{cr}=v'<c$. So it never approximates to the speed of light.

\par\noindent {\bf Example 2. Possibility of superluminal motion?} Take the relativistic-contraction factor
(\ref{L.6.2}) with $\epsilon=+1$ and non-parallel electric and magnetic fields. As we have seen above, this implies
$c<v'$, where $v'$ is a solution of $\dot xg\dot x=0$. Consider the non-minimal interaction
\begin{eqnarray}\label{L.6.7}
f^\mu=\frac{e}{mc^2}F^\mu{}_\nu D' x^\nu+\tilde f^\mu, \quad \mbox{where} \quad \tilde f^\mu=-\tilde
k^2D'x^\alpha\partial^\mu(FF)_{\alpha\beta}D'x^\beta.
\end{eqnarray}
We assume homogeneous and non-stationary fields with growing tension
\begin{eqnarray}\label{L.6.8}
\partial^i{\bf E}=\partial^i{\bf B}=0, \quad \frac{d}{dt}|{\bf E}|>0, \quad \frac{d}{dt}|{\bf B}|>0,
\end{eqnarray}
then $\tilde f^i=0$, $\tilde f^0=-\frac{\tilde k^2}{c\dot xg\dot x}\dot x\frac{\partial}{\partial t}(FF)\dot x$. The
longitudinal acceleration reads
\begin{eqnarray}\label{L.6.9}
{\bf v} {\bf a}=a_1(v)+a_2(v)\equiv\frac{e({\bf v}{\bf E})}{mc^3}\sqrt{-\dot xg\dot
x}(c^2-{\bf v}^2)-\frac{\tilde k^2{\bf v}^2}{c^2}\dot x\frac{\partial}{\partial t}(FF)\dot x.
\end{eqnarray}
We have $a_1(c)=0$, while $a_2(c)$ is positive according to Eqs. (\ref{L.6.4}) and (\ref{L.6.8}). So the particle
overcomes the light barrier. In the region $c<v<v'$ we have $a_1(v)<0$ and $a_2(v)>0$, so the particle will continue to
accelerate up to critical velocity $v_{cr}$ determined by the equation $a_1+a_2=0$. If $a_2>|a_1|$ in the region, the
particle will accelerate up to the value $v_{cr}=v'$.  Above this velocity, the equation (\ref{L.6.9}) becomes
meaningless.

The toy examples show that critical speed in a manifestly relativistic and reparametrization-invariant theory does not
always coincide with the speed of light, if we assume the usual special-relativity definitions of time and distance. In
general case, we expect that ${\bf v}_{cr}$ is both field and spin-dependent quantity. In the next section we repeat
this analysis for more realistic case of a particle with spin.

\section{Spinning particle in an arbitrary electromagnetic field}\label{sec7}

\subsection{Lagrangian and Hamiltonian formulations}\label{sec7.1}

In the formulation (\ref{FF.1}), the vector model of spin admits interaction with an arbitrary electromagnetic field. To introduce
coupling of the position variable $x$ with electromagnetic field we add the minimal interaction $\frac{e}{c}A_\mu\dot
x^\mu$. As for spin, it couples with $A^\mu$ through the term
\begin{eqnarray}\label{m.2}
D\omega^\mu\equiv\dot\omega^\mu-\lambda\frac{e\mu}{c}F^{\mu\nu}\omega_\nu\,,
\end{eqnarray}
where the coupling constant $\mu$ is the magnetic moment. They are the only terms we have found compatible with
symmetries and constraints which should be presented in the theory. Adding these terms to the free theory (\ref{FF.1})
we obtain the action
\begin{eqnarray}\label{m.1}
S=\int d\tau\frac{1}{4\lambda}\left[\dot xN\dot x+D\omega ND\omega-\sqrt{\left[\dot xN\dot x+D\omega
ND\omega\right]^2-4(\dot xND\omega)^2}\right]- \cr \frac{\lambda}{2}(m^2c^2-\frac{\alpha}{\omega^2})+\frac{e}{c}A\dot x.
\qquad \qquad \qquad \qquad \qquad
\end{eqnarray}
In their work \cite{hanson1974}, Hanson and Regge analyzed whether the spin-tensor (\ref{FF.2}) interacts directly with an
electromagnetic field, and concluded on impossibility to construct the interaction in closed form. In our model an
electromagnetic field interacts with the part $\omega^\mu$ of the spin-tensor.

Let us construct Hamiltonian formulation of the model. The procedure which leads to the Hamiltonian turns out to be
very similar to that described in subsection \ref{sec4.1},  so we present only the final expression  (for the details,
see \cite{deriglazov2014Monster}). Conjugate momenta for $x^\mu$, $\omega^\mu$ and $\lambda$ are denoted as $p^\mu$,
$\pi^\mu$ and $p_\lambda$.  We use also the canonical momentum ${\cal P}^\mu\equiv p^\mu-\frac{e}{c}A^\mu$. Contrary to
$p^\mu$, the canonical momentum ${\cal P}^\mu$ is $U(1)$ gauge-invariant quantity. With these notation, we obtain the
Hamiltonian variational problem which is equivalent to (\ref{m.1})
\begin{eqnarray}\label{m2.1}
S_H=\int d\tau ~ p_\mu\dot x^\mu+\pi_\mu\dot \omega^\mu+p_\lambda\dot\lambda-\left[\frac{\lambda}{2}\left( P^2-\frac{e\mu}{2c}(FS)+(mc)^2
+ \pi^2 - \frac{\alpha}{\omega^2}\right)+\lambda_2 (\omega\pi) +\lambda_3
(P\omega)+\lambda_4  (P\pi)+\lambda_0p_\lambda\right].
\end{eqnarray}
The expression in square brackets represents the Hamiltonian. By $\lambda_2$, $\lambda_3$, $\lambda_4$ and $\lambda_0$
we denoted the Lagrangian multipliers, they are written in front of the corresponding primary constraints.
%
%
The fundamental Poisson brackets $\{x^\mu, p^\nu\}=\eta^{\mu\nu}$ and $\{\omega^\mu, \pi^\nu\}=\eta^{\mu\nu}$ imply
\begin{eqnarray}\label{m.12}
\{x^\mu, {\cal P}^\nu\}=\eta^{\mu\nu}, \quad \{{\cal P}^\mu, {\cal P}^\nu\}=\frac{e}{c}F^{\mu\nu},
\end{eqnarray}
\begin{eqnarray}\label{m.13}
\{S^{\mu\nu},S^{\alpha\beta}\}= 2(\eta^{\mu\alpha} S^{\nu\beta}-\eta^{\mu\beta} S^{\nu\alpha}-\eta^{\nu\alpha}
S^{\mu\beta} +\eta^{\nu\beta} S^{\mu\alpha})\,.
\end{eqnarray}
According to Eq. (\ref{m.13}) the spin-tensor is generator of Lorentz algebra $SO(1,3)$. As $\omega\pi$, $\omega^2$ and
$\pi^2$ are Lorentz-invariants, they have vanishing Poisson brackets with $S^{\mu\nu}$. To reveal the higher-stage
constraints we write the equations $\dot T_i=\{T_i, H\}=0$. The Dirac procedure stops on  third stage with the
following equations
\begin{eqnarray}
p_\lambda=0 ~ ~ \quad &\Rightarrow& \quad  T_1\equiv{\cal P}^2-\frac{e\mu}{2c}(FS)+m^2c^2+\pi^2-\frac{\alpha}{\omega^2}=0 \quad
\Rightarrow \quad \lambda_3C+\lambda_4D=0\,,\label{m.14.1} \\
T_2\equiv (\omega\pi)=0 ~ ~ \quad &\Rightarrow& \quad ~  T_5\equiv\pi^2-\frac{\alpha}{\omega^2}=0\,,\label{m.14.2} \\
T_3\equiv({\cal P}\omega)=0 ~ \quad  &\Rightarrow & \quad ~  \lambda_4=-\frac{2\lambda c}{e}aC\,,\label{m.14.3} \\
T_4\equiv({\cal P}\pi)=0 ~ \quad &\Rightarrow&\quad ~  \lambda_3=\frac{2\lambda c}{e}aD\,.\label{m.14.4} 
\end{eqnarray}
We have denoted
\begin{eqnarray}\label{m.15}
C=-\frac{e(\mu-1)}{c}(\omega F{\cal P})+\frac{e\mu}{4c}(\omega\partial)(FS), \qquad  D=-\frac{e(\mu-1)}{c}(\pi F{\cal
P})+\frac{e\mu}{4c}(\pi\partial)(FS).
\end{eqnarray}
Besides, here and below we will use the following notation. In the equation which relates velocity and canonical
momentum will appear the matrix $T^{\mu\nu}$
\begin{eqnarray}\label{uf4.71}
T^{\mu\nu}=\eta^{\mu\nu}-(\mu-1)a(SF)^{\mu\nu}, \qquad
a=\frac{-2e}{4m^2c^3-e(2\mu+1)(SF)}.
\end{eqnarray}
Using the identity $S^{\mu\alpha}F_{\alpha\beta}S^{\beta\nu}=-\frac{1}{2}(S^{\alpha\beta}F_{\alpha\beta})S^{\mu\nu}$ we
find the inverse matrix
\begin{eqnarray}\label{pp7}
\tilde T^{\mu\nu}=\eta^{\mu\nu}+(\mu-1)b(SF)^{\mu\nu}, \qquad 
b=\frac{2a}{2+(\mu-1)a(SF)}\equiv\frac{-2e}{4m^2c^3-3e\mu(SF)},
\end{eqnarray}
The vector $Z^\mu$ is defined by
\begin{eqnarray}\label{gmm12}
Z^\mu=\frac{b}{4c}S^{\mu\sigma}(\partial_\sigma F_{\alpha\beta})S^{\alpha\beta}\equiv
\frac{b}{4c}S^{\mu\sigma}\partial_\sigma(FS).
\end{eqnarray}
This vanishes for homogeneous field, $\partial F=0$.
The last equation from (\ref{m.14.1}) turns out to be a consequence of (\ref{m.14.3}) and (\ref{m.14.4}) and can be
omitted. Due to the secondary constraint $T_5$ appeared in (\ref{m.14.2}) we can replace the constraint $T_1$ on the equivalent one
\begin{eqnarray}\label{gmm120}
T_1\equiv{\cal P}^2-\frac{e\mu}{2c}(FS)+m^2c^2=0.
\end{eqnarray}
The Dirac procedure revealed two secondary constraints written in Eqs. (\ref{gmm120}) and
(\ref{m.14.2}), and fixed the Lagrangian multipliers $\lambda_3$ and $\lambda_4$, the latter can be substituted into the Hamiltonian.
The multipliers $\lambda_0$,
$\lambda_2$ and the auxiliary variable $\lambda$ have not been determined. $H$ vanishes on the complete constraint surface,
as it should be in a reparametrization-invariant theory.

We summarized the algebra of Poisson brackets between constraints in the Table \ref{tabular:monster-algebra1}.
\begin{table}
\begin{center}
\begin{tabular}{c|c|c|c|c|c}
                              & $\qquad T_1 \qquad$  & $T_5$                   & $T_2$                 & $T_3$    & $T_4$     \\  \hline  \hline
$T_1=\mathcal{P}^2- $         & 0             & 0             & 0                                   & -2C   & -2D     \\
$\frac{\mu e}{2c}(FS)+m^2c^2$
                              &               &               &           &                           &             \\
\hline
$T_5=\pi^2-\frac{\alpha}{\omega^2}$               & 0             & 0                & $-2T_5\approx 0$          & $-2T_4\approx 0$   & $\frac{2\alpha}{(\omega^2)^2}T_3\approx 0$\\
& ${}$ &      &           &
&        \\
\hline
$T_2=\omega\pi$               & 0     & $2T_5\approx 0$       &0     & $-T_3\approx 0$     & $T_4\approx 0$\\
& ${}$ &      &           &
&        \\
\hline
$T_3={\cal P}\omega$       &$2C$  &$2T_4\approx 0$       &$T_3\approx 0$            & 0       & $T_1+\frac{e}{2ca}$  \\
&               &               &           &                                                                               &$\approx \frac{e}{2ca}$\\
\hline
$T_4={\cal P}\pi$          & $2D$      &$-\frac{2\alpha}{(\omega^2)^2}T_3\approx 0$           &$-T_4\approx 0$          &$-T_1-\frac{e}{2ca}$  &0\\

&    &       &                                                                                                                                      &$\approx -\frac{e}{2ca}$  &\\
\hline
\end{tabular}
\end{center}
\caption{Algebra of constraints.} \label{tabular:monster-algebra1}
\end{table}
The constraints $p_\lambda$, $T_1$, $T_2$ and $T_5$ form the first-class subset, while $T_3$ and $T_4$ represent a pair
of second class.  The presence of two primary first-class constraints $p_\lambda$ and $T_2$ is in correspondence with
the fact that two lagrangian multipliers remain undetermined within the Dirac procedure.

The evolution of the basic variables obtained according the standard rule $\dot z=\{z, H\}$ (equivalently, we can look
for the extremum of the variational problem (\ref{m2.1})). The equations read
\begin{eqnarray}\label{uf4.5}
\dot x^\mu=\lambda(T^{\mu}{}_\nu{\cal P}^\nu+\frac{\mu ca}{b}Z^\mu), \quad \qquad \dot{\cal P}^\mu=\frac{e}{c}(F\dot x)^\mu+\lambda\frac{\mu
e}{4c}\partial^\mu(FS),
\end{eqnarray}
\begin{eqnarray}\label{uf4.6}
\dot\omega^\mu=\lambda\frac{e\mu}{c}(F\omega)^\mu-\lambda\frac{2caC}{e}{\cal P}^\mu+\pi^\mu+\lambda_2\omega^\mu, \cr
\dot\pi^\mu=\lambda\frac{e\mu}{c}(F\pi)^\mu-\lambda\frac{2caD}{e}{\cal
P}^\mu-\frac{\alpha}{(\omega^2)^2}\omega^\mu-\lambda_52\pi^\mu,
\end{eqnarray}

The ambiguity due to the variables $\lambda$ and $\lambda_2$ means that the  interacting theory preserves both
reparametrization and spin-plane symmetries of the free theory. As a consequence, all the basic variables have
ambiguous evolution. $x^\mu$ and ${\cal P}^\mu$ have one-parametric ambiguity due to $\lambda$ while $\omega$ and $\pi$
have two-parametric ambiguity due to $\lambda$ and $\lambda_2$. The variables with ambiguous dynamics do not represent
observable quantities, so we look for the variables that can be candidates for observable quantities. We note that
(\ref{uf4.6}) imply an equation for $S^{\mu\nu}$ which does not contain $\lambda_2$
\begin{eqnarray}\label{uf4.9}
\dot S^{\mu\nu}&=& \lambda\frac{e\mu}{c}(FS)^{[\mu\nu]}+2{\cal P}^{[\mu}\dot x^{\nu]}\,.
\end{eqnarray}
This proves that the spin-tensor is invariant under local spin-plane symmetry. The remaining ambiguity due to $\lambda$
contained in Eqs. (\ref{uf4.5}) and (\ref{uf4.9}) is related with reparametrization invariance and disappears when we
work with physical dynamical variables $x^i(t)$. Thus we will work with $x^\mu$, ${\cal P}^\mu$ and $S^{\mu\nu}$.

The term $\frac{\alpha}{2\omega^2}$ in the Lagrangian (\ref{m.1}) provides the constraint $T_5$ which can be written as follows:
$\omega^2\pi^2=\alpha$. Together with $\omega\pi=0$, this implies fixed value of spin
\begin{eqnarray}\label{uf4.12.1}
S^{\mu\nu}S_{\mu\nu}=8(\omega^2\pi^2-(\omega\pi)^2)=8\alpha ,
\end{eqnarray}
for any solution to the equations of motion. The constraints $\omega{\cal P}=\pi{\cal P}=0$ imply the Pirani
condition for the spin-tensor (\ref{FF.2})
\begin{eqnarray}\label{uf4.12.2}
S^{\mu\nu}{\cal P_\nu}=0\,.
\end{eqnarray}
The equations (\ref{uf4.12.1}) and (\ref{uf4.12.2}) imply that only two components of spin-tensor are independent, as
it should be for spin one-half particle.

Equations (\ref{uf4.5}) and (\ref{uf4.9}), together with (\ref{uf4.12.1}) and (\ref{uf4.12.2}), form  a closed system
which determines evolution of a spinning particle.

The quantities $x^\mu$, $P^\mu$ and $S^{\mu\nu}$, being invariant under spin-plane symmetry, have vanishing brackets
with the corresponding first-class constraints $T_2$ and $T_5$. So, obtaining equations for these quantities, we can
omit the corresponding terms in the Hamiltonian (\ref{m2.1}). Further, we can construct the Dirac bracket for the
second-class pair $T_3$ and $T_4$. Since the Dirac bracket of a second-class constraint with any quantity vanishes, we
can now omit $T_3$ and $T_4$ from (\ref{m2.1}). Then the relativistic Hamiltonian acquires an expected form
\begin{eqnarray}\label{uf4.12}
H=\frac{\lambda}{2}\left({\cal P}^2-\frac{e\mu}{2c}(FS)+m^2c^2\right).
\end{eqnarray}
The equations (\ref{uf4.5}) and (\ref{uf4.9}) follow from this $H$ with use of Dirac bracket, $\dot z=\{z, H\}_{DB}$.

We can exclude the momenta ${\cal P}$ and the auxiliary variable $\lambda$ from the equations of motion. This yields
second-order equation for the particle's position. To achieve this, we solve the first equation from (\ref{uf4.5}) with
respect to ${\cal P}$ and use the identities $(SFZ)^\mu=-\frac12 (SF)Z^\mu$, $\tilde T^\mu{}_\nu
Z^\nu=\frac{b}{a}Z^\mu$, this gives
${\cal P}^\mu=\frac{1}{\lambda}\tilde T^\mu{}_\nu\dot x^\nu-\mu cZ^\mu$.
Then the Pirani condition reads $\frac{1}{\lambda}(S\tilde T\dot x)^\mu=\mu c(SZ)^\mu$. Using this equality, ${\cal
P}^2$ can be presented as ${\cal P}^2=\frac{1}{\lambda^2}(\dot x G\dot x)+\mu^2c^2Z^2$, where appeared the symmetric
matrix
\begin{eqnarray}\label{L.13}
G_{\mu\nu}=(\tilde T^T\tilde T)_{\mu\nu}=[\eta+b(\mu-1)(SF+FS)+b^2(\mu-1)^2FSSF]_{\mu\nu}.
\end{eqnarray}
The matrix $G$ is composed from the Minkowsky metric $\eta_{\mu\nu}$ plus (spin and field-dependent) contribution,
$G_{\mu\nu}=\eta_{\mu\nu}+h_{\mu\nu}(S)$. So we call $G$ the effective metric produced along the world-line by
interaction of spin with electromagnetic field. We substite ${\cal P}^2$ into the constraint (\ref{gmm120}), this gives
expression for $\lambda$
\begin{eqnarray}\label{L.13.1}
\lambda=\frac{\sqrt{-\dot x G\dot x}}{m_rc}, \qquad  m_r^2=m^2-\frac{\mu e}{2c^3}(FS)-\mu^2 Z^2\,.
\end{eqnarray}
This shows that the presence of $\lambda$ in Eq. (\ref{m.2}) implies highly non-linear interaction of spinning particle
with electromagnetic field. The final expression of canonical momentum through velocity is
\begin{eqnarray}\label{L.13.2}
{\cal P}^\mu=\frac{m_rc}{\sqrt{-\dot x G\dot x}}\tilde T^\mu{}_\nu\dot x^\nu-\mu cZ^\mu.
\end{eqnarray}
Using (\ref{L.13.1}) and (\ref{L.13.2}), we exclude ${\cal P}^\mu$ and $\lambda$ from the Hamiltonian equations
(\ref{uf4.5}), (\ref{uf4.9}) and (\ref{uf4.12.2}). This gives closed system of Lagrangian equations for the set $x, S$.
We have the dynamical equations
\begin{eqnarray}\label{FF.6}
D\left[m_r(\tilde T Dx)^\mu\right]=\frac{e}{c^ 2}(FDx)^\mu+ \frac{\mu
e}{4m_rc^3}\partial^\mu(SF)+\mu DZ^\mu\,,
\end{eqnarray}
\begin{eqnarray}\label{FF.7}
D S^{\mu\nu}=\frac{e\mu}{m_rc^2}(FS)^{[\mu\nu]}-2bm_rc(\mu-1)Dx^{[\mu}(SFDx)^{\nu]}+2\mu cD x^{[\mu}Z^{\nu]}\,,
\end{eqnarray}
the Lagrangian counterpart of Pirani condition
\begin{eqnarray}\label{FF.8}
S^{\mu\nu}[(\tilde T\dot x)_\nu-\frac{\mu\sqrt{-\dot xG\dot x}}{m_r}Z_\nu]=0,
\end{eqnarray}
as well as to the value-of-spin condition, $S^{\mu\nu}S_{\mu\nu}=8\alpha$. In the approximation $O^3(S, F, \partial F)$
and when $\mu=1$ they coincide with Frenkel equations, see \cite{DPM3}.

Eq. (\ref{FF.6}) shows how spin modifies the Lorentz-force equation (\ref{gm7.2}). In general case, the Lorentz force is modified due to the presence of (time-dependent) radiation mass $m_r$ (\ref{L.13.1}), the tetrad field $\tilde T$, the effective metric $G$ and due to two extra-terms on right hand side of (\ref{FF.6}).  Contribution of anomalous magnetic moment $\mu\ne 1$ to the
difference between $\dot x^\mu$ and ${\cal P}^\mu$ in (\ref{L.13.2}) is proportional to $\frac{J}{c^3}\sim\frac{\hbar}{c^3}$, while the
term with a gradient of field is proportional to $\frac{J^2}{c^3}\sim\frac{\hbar^2}{c^3}$.

Consider the homogeneous field,
\begin{eqnarray}\label{FF.8.1}
\partial_\alpha F^{\mu\nu}=0, \qquad Z^\mu=0.
\end{eqnarray}
Then contraction of (\ref{FF.8}) with $F_{\mu\nu}$ yields $(SF)\dot{}=0$, that is $S^{\mu\nu}F_{\mu\nu}$ turns out to be the
conserved quantity. This implies $\dot m_r=\dot a=\dot b=0$. Hence the Lorentz force is modified due to the presence of time-independent radiation mass $m_r$, the tetrad field $\tilde T$ and the effective metric $G$.

Consider the ``classical"  value of magnetic moment $\mu=1$. Then $\tilde T^{\mu\nu}=\eta^{\mu\nu}$ and $G_{\mu\nu}=\eta_{\mu\nu}$. The Lorentz force is modified due to the presence of time-dependent radiation mass $m_r$, and two extra-terms on right hand side of (\ref{FF.6}).

Let us specify the equation for spin precession to the case of uniform and stationary field, supposing also $\mu=1$ and
taking physical time as the parameter, $\tau=t$. Then (\ref{FF.8}) reduces to the Frenkel condition, $S^{\mu\nu}\dot
x_\nu=0$, while (\ref{FF.7}) reads $\dot S^{\mu\nu}=\frac{e\sqrt{-\dot x^2}}{m_rc^2}(FS)^{[\mu\nu]}$. We decompose
spin-tensor on electric dipole moment $\vec D$ and Frenkel spin-vector $\vec S$ according to (\ref{FF.2}), then $\vec
D=-\frac{2}{c}\vec S\times\vec v$, while precession of $\vec S$ is given by
\begin{eqnarray}\label{L.14.5}
\frac{d\vec S}{dt}=\frac{e\sqrt{c^2-\vec v^2}}{m_rc^3}\left[-\vec E\times(\vec v\times\vec S)+c\vec S\times\vec
B\right].
\end{eqnarray}

\subsection{Ultra-relativistic limit within the usual special-relativity notions}\label{sec7.2}

After identification $\theta_{\mu\nu}\equiv R_{\mu\nu\alpha\beta}S^{\alpha\beta}\sim F_{\mu\nu}$ and $\kappa\sim\mu$,
equations of motion in electromagnetic and in gravitational fields acquire the similar structure. Equations
(\ref{uf4.5}) and (\ref{uf4.9}) can be compared with (\ref{gmx-2})-(\ref{gmJ.3-2}) and (\ref{FF.6})-(\ref{FF.7}) with
(\ref{gml1}), (\ref{gml2}). In particular, in the Lagrangian equations with anomalous magnetic moment ($\mu\ne 1$) in
Minkowski space also appeared the effective metric (\ref{L.13}). So we need to examine the ultra-relativistic limit. In
this section we do this under the usual special-relativity notions, that is we suppose that the particle probes the
three-dimensional geometry (\ref{L.1.1}). We show that the critical speed turns out to be different from the speed of
light while an acceleration, contrary to subsection \ref{sec4.4}, vanishes in ultra-relativistic limit. It will be
sufficient to estimate the  acceleration in the uniform and stationary field (\ref{FF.8.1}). We take $\tau=t$ in
equations (\ref{FF.6})-(\ref{FF.8}) and compute the time derivative on l. h. s. of equations (\ref{FF.6}) with $\mu=1,
2, 3$. Then the equations read
\begin{eqnarray}\label{FF.15.0}
a^i-\frac{v^i}{2(-vGv)}\frac{d}{dt}(-vGv)=T^i{}_\nu\left[\frac{e\sqrt{-vGv}}{m_rc^ 2}(Fv)^i-\frac{d}{dt}\tilde T^\nu{}_\alpha v^\alpha\right]\,,
\end{eqnarray}
\begin{eqnarray}\label{FF.16}
\frac{d}{dt}S^{\mu\nu}=\frac{e\mu\sqrt{-vGv}}{m_rc^2}(FS)^{[\mu\nu]}- \frac{2bm_rc(\mu-1)}{\sqrt{-vGv}}v^{[\mu}(SFv)^{\nu]}\,,
\end{eqnarray}
\begin{eqnarray}\label{FF.17}
(Sv)^\mu+b(\mu-1)(SSFv)^\mu=0,
\end{eqnarray}
where $v^\mu=(c, {\bf v})$.
Eqs. (\ref{FF.17}) and (\ref{L.13}) imply
\begin{eqnarray}\label{FF.17.1}
-vGv=-v\tilde Tv=c^2-{\bf v}^2-(\mu-1)b(vSFv).
\end{eqnarray}
We compute the time-derivatives in Eq. (\ref{FF.15.0})
\begin{eqnarray}\label{FF.15.1}
\frac{d}{dt}(-vGv)=-2({\bf v}{\bf a})-(\mu-1)b\left\{ [v(FS+SF)]_ia^i+\frac{e\mu\sqrt{-vGv}}{m_rc^2}[(vFFSv)+(vFSFv)]-\right. \cr
\left. \frac{2bm_rc(\mu-1)}{\sqrt{-vgv}}[v^2(vFSFv)-(vSFv)(vFv)]\right\},
\end{eqnarray}
\begin{eqnarray}\label{FF.15.2}
-T^i{}_\nu\frac{d}{dt}\tilde T^\nu{}_\alpha v^\alpha=-\frac{e\sqrt{-vGv}}{m_rc^2}\left\{\mu(\mu-1)b(FSFv)^i-
\mu(\mu-1)a(SFFv)^i-\mu(\mu-1)^2ab(SFFSFv)^i\right\}+ \cr
\frac{2bm_rc(\mu-1)}{\sqrt{-vGv}}T^i{}_\nu[v^\nu(vFSFv)-(SFv)^\nu(vFv)].
\end{eqnarray}
We note that all the potentially divergent terms (two last terms in (\ref{FF.15.1}) and in (\ref{FF.15.2})), arising due
to the contribution from $\dot S\sim\frac{1}{\sqrt{-vGv}}$, disappear on the symmetry grounds. We substitute non
vanishing terms into (\ref{FF.15.0}) obtaining the expression
\begin{eqnarray}\label{FF.15}
M^i{}_ja^j= \frac{e\sqrt{-vGv}}{m_rc^2}\left\{(Fv)^i-\mu(\mu-1)b(FSFv)^i+(\mu-1)^2a(SFF[\eta+\mu bSF]v)^i-
v^i\frac{\mu(\mu-1)b}{2(-vGv)}(vFFSv)\right\},
\end{eqnarray}
where the matrix
\begin{eqnarray}\label{FF.15.3}
M^i{}_j=\delta^i{}_j+\frac{v^iv^\mu\Omega_{\mu j}}{2(-vGv)},
\quad \mbox{with} \quad  \Omega_{\mu j}=2\delta_{\mu j}+(\mu-1)b(FS+SF)_{\mu j},
\end{eqnarray}
has the inverse
\begin{eqnarray}\label{FF.15.4}
\tilde M^i{}_j=\delta^i{}_j-\frac{v^iv^\mu\Omega_{\mu j}}{2c^2-(\mu-1)bv^\mu(FS+SF)_{\mu 0}v^0},
\end{eqnarray}
with the property
\begin{eqnarray}\label{FF.15.5}
\tilde M^i{}_jv^j=v^i\frac{2(-vGv)}{2c^2-(\mu-1)bv^\mu(FS+SF)_{\mu 0}v^0}.
\end{eqnarray}
Applying the inverse matrix we obtain the acceleration
\begin{eqnarray}\label{FF.15.6}
a^i= \frac{e\sqrt{-vGv}}{m_rc^2}\left\{\tilde M^i{}_j[(Fv)^j-\mu(\mu-1)b(FSFv)^j+(\mu-1)^2a(SFF[\eta+\mu bSF]v)^j]-\right. \cr
\left. v^i\frac{\mu(\mu-1)b(vFFSv)}{2c^2-(\mu-1)bv^\mu(FS+SF)_{\mu 0}v^0}\right\}.
\end{eqnarray}
For the particle with non anomalous magnetic moment ($\mu=1$), the right hand side reduces to the Lorentz force, so the
expression in braces is certainly non vanishing in the ultra-relativistic limit. Thus the acceleration vanishes only
when $v\rightarrow v_{cr}$, where the critical velocity is determined by the equation $vGv=0$.

Let us estimate the critical velocity.  Using the consequence $(\dot xSF\dot x)=-b(\mu-1)(\dot xFSSF\dot x)$ of the Pirani
condition, and the expression $S^{\mu}{}_\alpha
S^{\alpha\nu}=-4\left[\pi^2\omega^\mu\omega^\nu+\omega^2\pi^\mu\pi^\nu\right]$, we write
\begin{eqnarray}\label{FFF.6.1}
-(\dot xG\dot x)=c^2-{\bf v}^2+4b^2(\mu-1)^2\left[\pi^2(\omega F\dot
x)^2+\omega^2(\pi F\dot x)^2\right].
\end{eqnarray}
As $\pi$ and $\omega$ are space-like vectors (see the discussion below Eq. (\ref{T0})), the last term is non-negative,
so $v_{cr}\ge c$. We show that generally this term is nonvanishing function of velocity, then $v_{cr}> c$. Assume the
contrary, that this term vanishes at some velocity, then
\begin{eqnarray}\label{FFF.7.1}
\omega F\dot x=-\omega^0({\bf E}{\bf v})+({\boldsymbol{\omega}}, c{\bf E}+{\bf v}\times{\bf B})=0\,, \cr \pi F\dot
x=-\pi^0({\bf E}{\bf v})+({\boldsymbol{\pi}}, c{\bf E}+{\bf v}\times{\bf B})=0\,.
\end{eqnarray}
This implies $c({\bf D}{\bf E})+({\bf D}, {\bf v}\times{\bf B})=0$. Consider the case ${\bf B}=0$, then it should be
$({\bf D}{\bf E})=0$. On other hand, for the homogeneous field the quantity $S^{\mu\nu}F_{\mu\nu}=2\left[({\bf D}{\bf
E})+2({\bf S}{\bf B})\right]=2({\bf D}{\bf E})$ is a constant of motion. Hence we can take the initial
conditions for spin such, that $({\bf D}{\bf E})\ne 0$ at any instant, this implies $v_{cr}>c$.

\subsection{Ultra-relativistic limit within the geometry determined by effective metric}\label{sec7.3}

According to the previous section, if we insist to preserve the usual special-relativity definitions of time and
distance (\ref{L.1.1}), the speed of light does not represent special point of the equation for trajectory.
Acceleration of the particle with anomalous magnetic moment generally vanishes at the speed slightly higher than the speed of
light. Hence we arrive at a rather surprising result that speed of light does not represent maximum velocity of the
manifestly relativistic equation (\ref{FF.15}). This state of affairs is unsatisfactory because the Lorentz
transformations have no sense above $c$, so two observers with relative velocity $c<v<v_{cr}$ will not be able to
compare results of their measurements.

To keep the condition $v_{cr}=c$, we use formal similarity of the matrix $G$ appeared in (\ref{L.13}) with space-time
metric. Then we can follow the general-relativity prescription of section \ref{sec2} to define time and distance in the
presence of electromagnetic field. That is we use $G$ of Eq. (\ref{L.13}) to define the three-dimensional geometry
(\ref{La.3.0})-(\ref{La.5}). The effective metric depends on $x^i$ via the field strength $F(x^0, x^i)$, and on $x^0$
via the field strength as well as via the spin-tensor $S(x^0)$. So the effective metric is time-dependent even in
stationary electromagnetic field. With these definitions we have, by construction, $-\dot x G \dot x=
(\frac{dt}{dx^0})^2(c^2 - ({\bf v}\gamma{\bf v}))$, so the critical speed coincides with the speed of light. The
intervals of time and distance are given now by Eq. (\ref{La.3.0}) and (\ref{La.3}), they slightly differ from those in
empty space.

In the present case, the expression for three-acceleration can be obtained in closed form in an arbitrary
electromagnetic field. We present Eq. (\ref{FF.6}) in the form (\ref{La.20.7})
\begin{eqnarray}\label{FF.16}
DDx^\mu={\mathcal F}^\mu=-Dx^\mu\frac{Dm_r(S)}{m_r}-T^\mu{}_\nu D\tilde
T^\nu{}_\alpha(S)Dx^\alpha+T^\mu{}_\nu\left\{\frac{e}{m_rc^ 2}(FDx)^\nu+ \frac{\mu
e}{4m^2_rc^3}\partial^\nu(SF)+\frac{\mu}{m_r} DZ^\nu\right\}\,.
\end{eqnarray}
Then the acceleration is given by (\ref{La.20.9}). The first two terms on right hand side of (\ref{FF.16}) give
potentially divergent contributions arising from the piece $\dot S\sim\frac{1}{\sqrt{c^2-{\bf v}\gamma{\bf v}}}$ of Eq.
(\ref{FF.7}). In the previous section we have seen that the dangerous contribution contained in the second term
disappears. To analyse the first term, we substitute ${\mathcal F}^i$ from (\ref{FF.16}) into (\ref{La.20.9}). With use
the property $\tilde M^i{}_jv^j=v^i\frac{c^2-{\bf v}\gamma{\bf v}}{c^2}$, we obtain the acceleration
\begin{eqnarray}\label{FF.170}
a^i=(c^2-{\bf v}\gamma{\bf v})\left[-v^i\frac{\dot m_r}{m_rc^2}-\frac{\tilde M^i{}_jT^j{}_\nu\dot{\tilde
T}^\nu{}_\alpha v^\alpha}{c^2-{\bf v}\gamma{\bf v}}+\right. \qquad \qquad \qquad \quad \cr \left. \tilde
M^i{}_jT^j{}_\nu\left\{\frac{e}{m_rc^2\sqrt{c^2-{\bf v}\gamma{\bf v}}}(Fv)^\nu+ \frac{\mu
e}{4m^2_rc^3}\partial^\nu(SF)+\frac{\mu}{m_r\sqrt{c^2-{\bf v}\gamma{\bf v}}} \dot Z^\nu\right\}\right]+ \cr \tilde
M^i{}_j\tilde\Gamma^j{}_{kl}(\gamma)v^kv^l+\frac12\left(\frac{dt}{dx^0}\right)^{-1}\left[({\bf
v}\partial_0\gamma\gamma^{-1})^i-\frac{v^i}{c^2}({\bf v}\partial_0\gamma{\bf v})\right], \qquad \quad
\end{eqnarray}
so the divergency due to $\dot m_r\sim\frac{1}{\sqrt{c^2-{\bf v}\gamma{\bf v}}}$ is cancelled by the factor in front of
this term. In the result, the acceleration is finite as $v\rightarrow c$. Besides, taking into account the property
$({\bf v}\gamma)_i\tilde M^i{}_j=({\bf v}\gamma)_j\frac{c^2-{\bf v}\gamma{\bf v}}{c^2}$, we conclude that the
longitudinal acceleration (\ref{La.20.10})
\begin{eqnarray}\label{FF.18}
{\bf v}\gamma{\bf a}=\frac{(c^2-{\bf v}\gamma{\bf v})^2}{c^2}({\bf v}\gamma{\boldsymbol{\mathcal F}})+\frac{c^2-{\bf v}\gamma{\bf v}}{c^2}
\left[({\bf v}\gamma)_i\tilde\Gamma^i{}_{kl}(\gamma)v^kv^l+\frac12\left(\frac{dt}{dx^0}\right)^{-1}({\bf v}\partial_0\gamma{\bf v})\right],
\end{eqnarray}
vanishes in this limit.


\section{Conclusion}\label{sec8}

In this work we have studied behavior of ultra-relativistic spinning particle in external fields. To construct
interaction of spin with external fields and to analyze its influence on the trajectory of the particle, we used the
vector model of spin. Minimal interaction  with gravity was formulated starting from the Lagrangian variational problem
without auxiliary variables (\ref{L-curved}). The non-minimal interaction with gravity through the gravimagnetic moment
$\kappa$ \cite{Khriplovich1989} has been achieved in the Lagrangian with one auxiliary variable (\ref{acc-11}).

 The variational problems imply the fixed value of spin (\ref{condition2}), that is, they correspond to an elementary spin
one-half particle. The vector model also allowed us to construct the Lagrangian action (\ref{aaa1}) with unfixed spin
and with a mass-spin trajectory constraint, that is with the properties of Hanson-Regge relativistic top
\cite{hanson1974}. In this model appeared the fundamental length scale and spin has four physical degrees of freedom.
At last, interaction of spinning particle with magnetic moment $\mu$ with an arbitrary electromagnetic field was
achieved in the Lagrangian action with one auxiliary variable (\ref{m.1}).
Equations of motion of minimally interacting spinning particle (that is with $\kappa=0$) have been identified with
Mathisson-Papapetrou-Tulczyjew-Dixon equations. They are widely used in the current literature for description of
rotating bodies in general relativity. To study the class of trajectories of a body with fixed integrals of motion
$\sqrt{-P^2}=k$ and $S^2=\beta$, we can use our spinning particle with $m=\frac{k}{c}$ and $\alpha=\frac{\beta}{8}$.

To study our general-covariant equations in the laboratory frame, we used the Landau-Lifshitz $1+3$\,-splitting formalism of four-dimensional pseudo Riemann space, where the basic structure is a congruence of one-dimensional timelike curves identified with worldlines of the laboratory clocks. This formalism allows one to determine the time interval, distance and then velocity between two infinitesimally closed points $x^\mu$ and $x^\mu+\delta x^\mu$ of the particle's worldline. The basaic requirement for definition of the three-dimensional quantities is that speed of light should be a coordinate-independent notion. Due to the decomposition of spacetime into time $+$ space, one manipulates only with time-varying vector and  tensor fields.  In the resulting three-dimensional geometry with Riemannian scalar product, we asked on the notion of a constant vector field. We have suggested the notion (\ref{La.8.4}) which follows from the geometric requirement that scalar product of constant fields does not depends on the point where it was computed. For the vector field of velocity, its deviation from the constant field has given us the acceleration (\ref{La.8.5}). Then we showed that the definition adopted is consistent with the basic principle of general relativity: massive spinless particle, propagating in a gravitational field along a four-dimensional geodesic, can not exceed the speed of light. With this definition at hands, we analysed ultra relativistic behavior of the spinning particle in external fields.

Evolution of the fast MPTD-particle in the laboratory frame was studied on the base of Lagrangian equations
(\ref{byM13}) and (\ref{motionJ-5}). In these equations we observed the emergence of the effective metric
(\ref{G-metric}) instead of the original one. We have examined the two metrics as candidates for construction of
three-dimensional space-time geometry (\ref{La.3.0})-(\ref{La.5}) probed by the particle. In both cases the
MPTD-equations have unsatisfactory behavior in the ultra-relativistic limit. In particular, the three-dimensional
acceleration (\ref{La.8.5}) increases with velocity and becomes infinite in the limit.

Further, we showed that spinning particle with $\kappa=1$ is free of the problems detected in MPTD-equations. For this
value of gravimagnetic moment the effective metric does not appeared and the three-dimensional geometry should be
defined, unambiguously, with respect to the original metric. Critical velocity of the theory coincides with the speed
of light and three-dimensional acceleration vanishes as $v\rightarrow c$. So the spinning particle with gravimagnetic
moment $\kappa=1$ seems more promising candidate for the description of a relativistic rotating body in general
relativity. An interesting property of the resulting equations is that spin ceases to affect the trajectory in
ultra-relativistic limit: the trajectory of spinning particle becomes more and more close to that of spinless particle
as $v\rightarrow c$. Besides, the spin precesses with finite angular velocity in this limit.

Equations in electromagnetic and in gravitational fields become very similar after the identification $\mu\sim\kappa$
and $R_{\mu\nu\alpha\beta}S^{\alpha\beta}\sim F_{\mu\nu}$. In particular, interaction of spin with electromagnetic
field in Minkowski space also produces the effective metric (\ref{L.13}) for the particle with anomalous magnetic
moment $\mu\ne 1$. If we insist on the usual special-relativity notions of time and distance, the critical speed turns
out more than the speed of light. To preserve the equality $v_{cr}=c$, we are forced to assume that particle in
electromagnetic field probes the three-dimensional geometry determined with respect to the effective metric instead of
the Minkowski metric. In the result, we have rather unusual picture of the Universe filled with spinning matter. Since
$G$ depends on spin, in this picture there is no unique space-time manifold for the Universe of spinning particles:
each particle will probe his own three-dimensional geometry. In principle this could be an observable effect. With the
effective metric (\ref{L.13}), the equation (\ref{La.3.0}) implies that the time of life of muon in electromagnetic
field and in empty space should be different.

\section*{Acknowledgments}
The work of AAD has been supported by the Brazilian foundations CNPq (Conselho Nacional de Desenvolvimento Cient\'ifico e
Tecnol\'ogico - Brasil) and FAPEMIG (Fundac\~ao de Amparo \'a Pesquisa do Estado de Minas Gerais - Brasil). WGR thanks CAPES
for the financial support (Programm PNPD/2011).

\end{document}